\documentclass[%
 reprint,
nofootinbib,
 amsmath,amssymb,
 aps,
floatfix,
]{revtex4-1}

\usepackage{graphicx}
\usepackage{dcolumn}
\usepackage{bm}
\usepackage[caption=false]{subfig}

\begin{document}


\title{Novel constraints on non-cold (non-thermal) Dark Matter from Lyman-$\alpha$ forest data}

\author{Riccardo Murgia}
\email{riccardo.murgia@sissa.it}
\affiliation{SISSA, Via Bonomea 265, 34136 Trieste, Italy}
\affiliation{INFN, Sezione di Trieste, Via Bonomea 265, 34136 Trieste, Italy} 

\author{Vid Ir\v{s}i\v{c}}%
\email{irsic@uw.edu}
\affiliation{University of Washington, Department of Astronomy,
3910 15th Ave NE, WA 98195-1580 Seattle, USA}

\author{Matteo Viel}
\email{viel@sissa.it}
\affiliation{SISSA, Via Bonomea 265, 34136 Trieste, Italy}
\affiliation{INFN, Sezione di Trieste, Via Bonomea 265, 34136 Trieste, Italy} 
\affiliation{INAF/OATS, Osservatorio Astronomico di Trieste, via Tiepolo 11, I-34143 Trieste, Italy}


\begin{abstract}
In this paper we present an efficient method for constraining both thermal and non-thermal Dark Matter (DM) scenarios with the Lyman-$\alpha$ forest, based on a simple and flexible parametrisation capable to reproduce the small scale clustering signal of a large set of non-cold DM (nCDM) models.
We extract new limits on the fundamental DM properties, through an extensive analysis of the high resolution, high redshift data obtained by the MIKE/HIRES spectrographs.
By using a large suite of hydrodynamical simulations, we determine constraints on both astrophysical, cosmological, and nCDM parameters by performing a full Monte Carlo Markov Chain (MCMC) analysis. We obtain a marginalised upper limit on the largest possible scale at which a power suppression induced by nearly any nCDM scenario can occur,~i.e.~$\alpha<0.03~{\rm{Mpc}}/h$ (2$\sigma$ C.L.).
We explicitly describe how to test several of the most viable nCDM scenarios without the need to run any specific numerical simulations, due to the novel parametrisation proposed, and due to a new scheme that interpolates between the cosmological models explored.
The shape of the linear matter power spectrum for standard thermal warm DM models appear to be in mild tension ($\sim 2\sigma$ C.L.) with the data, compared to non-thermal scenarios.
We show that a DM fluid composed by both a warm (thermal) and a cold component is also in tension with the Lyman-$\alpha$ forest, at least for large $\alpha$ values.
This is the first study that allows to probe the linear small scale {\it shape} of the DM power spectrum for a large set of nCDM models. 

\end{abstract}

\maketitle


\section{\label{sec:intro}Introduction}
Dark Matter (DM) candidates are typically classified according to their velocity dispersion, which defines the so-called free-streaming length.
On scales smaller than their free-streaming length, density fluctuations are wiped out and gravitational clustering is suppressed. The velocity dispersion of Cold DM (CDM) candidates is by definition so small that the corresponding free-streaming length does not affect cosmological structure formation.

Assuming the standard CDM model, $N$-body simulations
predict too many dwarf galaxies within the Milky Way (MW) virial radius (\emph{missing satellite} problem~\cite{Klypin:1999uc,Moore:1999nt}) and too much DM in the innermost regions of galaxies (\emph{cusp-core} problem~\cite{2010AdAst2010E...5D}), with respect to the observations.
Moreover, the dynamical properties of the most massive MW satellites are not correctly predicted by simulations (\emph{too-big-to-fail} problem~\cite{2011MNRAS415L40B,2012MNRAS4221203B}). These inconsistencies, often denoted as the CDM \emph{small-scale crisis} (e.g.~\cite{bullock18} and references therein), may be relaxed either by baryon physics, currently difficult to be efficiently implemented in cosmological simulations~\cite{Okamoto:2008sn,Governato:2012fa}, or by
modifying the standard CDM framework, given that the fundamental nature of DM is still unknown~\cite{Schneider:2016ayw,Lapi:2013bxa,salucci13}.

Therefore, various non-cold DM (nCDM) scenarios predicting structure formation to be suppressed at small cosmological
scales, have been investigated as a viable solution for the small-scale crisis (e.g., sterile neutrinos~\cite{Adhikari:2016bei,Konig:2016dzg,devega}, ultra-light scalars~\cite{Hu:2000ke,Hui:2016ltb,takeshi}, mixed cold plus warm DM fluids~\cite{Schneider:2016ayw,Viel2005},
DM-Dark Radiation interaction models or Self-Interacting DM~\cite{Cyr-Racine:2015ihg,Vogelsberger:2015gpr}, DM coupled with Dark Energy~\cite{Kobayashi:2018nzh,Murgia:2016ccp}).

The suppression in the matter power spectrum induced by nCDM can be characterised by different strength and shape, depending 
on the fundamental nature of the DM candidate.
Up to now, lots of efforts have gone into examining the astrophysical consequences of thermal Warm DM (WDM) models, i.e.~candidates with a Fermi-Dirac or Bose-Einstein momentum distribution, which
implies a very specific shape of the small-scale suppression, only depending on the WDM particle mass~\cite{Viel:2013apy,Lapi:2015zea,Irsic:2017ixq}.
However, most of the aforementioned nCDM candidates, well motivated by theoretical particle physics, do not feature a thermal momentum distribution: this can result in non-trivial suppressions in their power spectra, not appropriately described
by the oversimplified thermal WDM case~\cite{Baur:2017stq}.
For these reasons, in Ref.~\cite{Murgia:2017lwo} we introduced a new, general parametrisation for the small-scale power suppression, which accurately covers all the most viable (non-thermal) nCDM scenarios, with the goal to provide a fully general modelling of the small-scale departures from the standard CDM model. Such parametrisation represents a direct link between DM model building and structure formation observations. This can be exploited to investigate, in a simple way, the astrophysical implications of different nCDM scenarios, focusing not only on the corresponding cut-off scale, but also on the peculiar features of the power spectra at small scales. Such investigation is therefore intriguing~{\emph{per~se}, even regardless of the CDM small-scale crisis.

In this paper, we present the first accurate astrophysical constraints on the general parametrisation discussed in Ref.~\cite{Murgia:2017lwo}, which are easily translatable to bounds on the fundamental nCDM properties.
They have been obtained through a comprehensive analysis of the Lyman-$\alpha$ forest~\cite{McQuinn:2015icp}, namely the absorption lines produced by the inhomogeneous distribution of
the intergalactic neutral hydrogen along different line of sights to distant quasars (QSOs)~\cite{Viel:2001hd}, which is an ideal tracer for the matter power spectrum at high redshifts ($2 \lesssim z \lesssim 5$) and small scales ($0.5~{\rm Mpc}/h, \lesssim  \lambda \lesssim 20~{\rm Mpc}/h$)~\cite{Viel:2013apy,Irsic:2017ixq}.

The paper is organised as follows: in Section~\ref{sec:param} we briefly summarise the novel parametrisation for the small-scale power suppression; in Section~\ref{sec:sims} we describe the suite of simulations that we have performed; in Section~\ref{sec:data} we present the data set that we have used; in Section~\ref{sec:method} we discuss the method that we have adopted for the analysis; in Section~\ref{sec:results} we discuss the results that we have obtained and their implications for the fundamental nature of DM; finally, in Section~\ref{sec:conclusions} we draw the conclusions and outline the future developments of this work.

\section{\label{sec:param}A new, general parametrisation}
\begin{figure}
\includegraphics[width=0.49\textwidth]{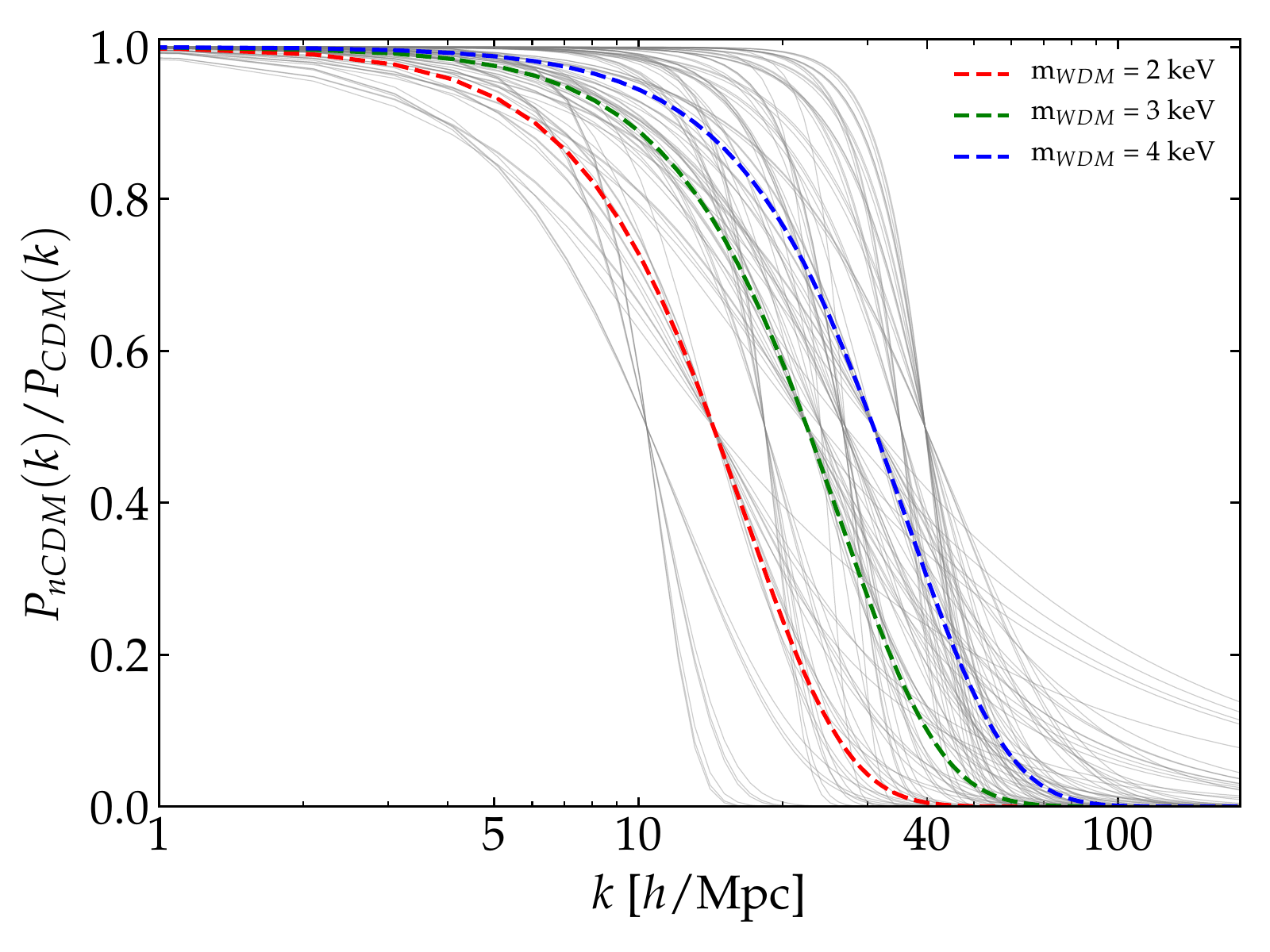}
\caption{Here we plot the squared transfer functions associated to the 109 $\{\alpha,\beta,\gamma\}$-combinations that we used for our analysis (grey solid lines), each of them corresponding to a different nCDM model (see also Table~\ref{tab:params}). We also plot the squared transfer functions corresponding to three thermal WDM models with masses 2, 3 and 4 keV (blue, green and red dashed lines, respectively).}
\label{fig:Tk}
\end{figure}

The small-scale suppression of the matter power spectrum $P(k)$, due to the existence of nCDM, is usually described by the transfer function $T(k)$, which is defined as follows:
\begin{equation}\label{eq:Tkdef}
 T^2(k) = \left[ \frac{P(k)_{\rm{nCDM}}}{P(k)_{\rm{CDM}}} \right],
\end{equation}
i.e.~as the square root of the ratio of the linear power spectrum in the presence of nCDM
with respect to that in the presence of CDM only, for fixed cosmological parameters.
For the particular case of thermal WDM, the transfer function is well approximated by the analytical fitting function~\cite{Bode2001}
\begin{equation}\label{eq:Viel}
 T(k) = [ 1 + (\alpha k)^{2\mu} ]^{-5/\mu},
\end{equation}
where $\alpha$ is the only free parameter and $\mu = 1.12$.
Therefore, bounds on the mass of the thermal WDM candidate are easily converted into constraints on $\alpha$,
through the following formula~\cite{Viel2005}:
\begin{equation}\label{eq:alphaold}
\begin{aligned}
 \alpha = 0.24 \left( \frac{m_x/T_x}{1~\rm{keV}/T_\nu} \right)^{-0.83} \left( \frac{\omega_x}{0.25(0.7)^2} \right)^{-0.16}{\rm{Mpc}}\\
 = 0.049 \left( \frac{m_x}{1~\rm{keV}} \right)^{-1.11} \left( \frac{\Omega_x}{0.25} \right)^{0.11} \left(\frac{h}{0.7}\right)^{1.22}h^{-1}\rm{Mpc}~,
\end{aligned}
\end{equation}
with $m_i$ being the mass, $T_i$ the temperature, $\Omega_i$ the abundance of the $i$-th species and $\omega_i \equiv \Omega_ih^2$. The index $i = x,\nu$ stands for WDM and active neutrinos, respectively.

Let us now introduce the half-mode scale, $k_{1/2}$, as the wave-number for which $T^2\equiv0.5$,
and define the following generalisation of Eq.~\eqref{eq:Viel}, which has been introduced and thoroughly discussed in Refs. \cite{Murgia:2017lwo,Murgia:2017cvj}:
\begin{equation}\label{eq:Tgen}
 T(k) = [ 1 + (\alpha k)^{\beta} ]^{\gamma},
\end{equation}
so that $k_{1/2}$ is a function of the three parameters $\alpha$, $\beta$ and $\gamma$,~i.e.
\begin{equation}\label{eq:k12}
 k_{1/2} = ((0.5)^{1/2\gamma}-1)^{1/\beta})\alpha^{-1}.
\end{equation}
Via Eqs.~\eqref{eq:Viel} and \eqref{eq:alphaold} we have a one-to-one correspondence between $m_x$ and
$\alpha$. On the other hand, through Eqs.~\eqref{eq:Tgen} and \eqref{eq:k12}, bounds on the DM mass are mapped to
3D surfaces in the $\{\alpha,\beta,\gamma\}$-space. In other words, given a value of $k_{1/2}$ which corresponds 
to a certain thermal WDM mass, Eq.~\eqref{eq:k12} allows to compute the corresponding surface in a 3D parameter space.

It is well established that thermal warm DM candidates with masses of the order of 3 keV can induce a suppression in 
the corresponding matter power spectra such that the CDM small-scale crisis vanishes or it is largely reduced~\cite{Lovell:2015psz,Lovell:2016nkp}.
Hence, it is compelling to investigate the volume of the $\{\alpha,\beta,\gamma\}$-space associated to thermal WDM masses roughly between 2 and 4~keV. This can be done by building a 3D grid in the parameter space which samples that volume, with each of the grid points unequivocally identified by a certain $\{\alpha,\beta,\gamma\}$-combination, corresponding to a different nCDM model.

In Fig.~\ref{fig:Tk} we plot 109 transfer functions, computed through Eq.~\eqref{eq:Tgen} and associated with the $\{\alpha,\beta,\gamma\}$-combinations reported in Table~\ref{tab:params}, which approximately bracket the aforementioned region of the parameter space.

Notice that the position of the half-mode scale $k_{1/2}$ is still primarily set by the value of $\alpha$, even in the new, general
parametrisation, while $\beta$ and $\gamma$ are responsible for the slope of the transfer functions before and after $k_{1/2}$, respectively.
$\beta$ has to be greater than zero in order to have physical transfer functions, since negative values for $\beta$ lead to transfer functions which increase with $k$ and reach 1 at small scales.
The larger is $\beta$, the flatter is the shape for $k < k_{1/2}$; the larger is $|\gamma|$, the steeper is the small-scale cut-off.

Fig.~\ref{fig:Tk} manifestly shows that the new fitting formula is flexible enough to disentangle even tiny differences in the shape of the power suppression and, thus, to discriminate between distinct (non-thermal) nCDM models with power spectra suppressed at very similar scales.
It has been already shown in Ref.~\cite{Murgia:2017lwo} that such versatility is extremely useful in order to provide a direct link between the fundamental particle nature of DM and the astrophysical observations, given that Eq.~\eqref{eq:Tgen} accurately reproduces the small-scale power suppression induced by all the most popular nCDM scenarios. The fitted transfer functions practically always lead to the same conclusion provided by the actual theoretical particle physics models (see Refs.~\cite{Murgia:2017lwo,Murgia:2017cvj} for more details).

\section{\label{sec:sims}Simulations}
Our analysis is based on a large suite of hydrodynamical simulations, performed with {\tt GADGET-3}, a modified version of the publicly available {\tt GADGET-2} code~\cite{Springel:2005mi,Springel:2000yr}.
As in Ref.~\cite{Irsic:2017ixq}, our reference model simulation has a box length of $20/h$ comoving Mpc with $2 \times 768^3$ gas and CDM particles (with gravitational
softening $1.04/h$ comoving kpc) in a flat $\Lambda$CDM universe
with cosmological parameters $\Omega_m = 0.301$, $\Omega_b = 0.0457$,
$n_s = 0.961$, $H_0 = 70.2$~km~s$^{-1}$~Mpc$^{-1}$ and $\sigma_8 = 0.829$~\cite{Ade:2015xua}.
Since the physical observable for Lyman-$\alpha$ forest experiments is the \emph{flux power spectrum} $P_{\rm{F}}(k,z)$, the goal of our set of simulations is to provide a reliable 
template of mock flux power spectra to be compared with observations.

Given that the flux power spectrum is affected both by astrophysical and cosmological parameters, it is important to properly take them into account and accurately quantify their impact in the likelihood. To this end, along the lines of Ref.~\cite{Irsic:2017ixq}, we have explored several values of the cosmological parameters $\sigma_8$,~i.e.~the
normalisation of the linear matter power spectrum, and $n_{\rm eff}$,~namely the slope of the matter power spectrum at the scale of Lyman-$\alpha$ forest (0.009 s/km). It is nowadays well established that varying these two parameters is sufficient to properly accounting for the effect on the matter power spectrum due to changes in its initial slope and amplitude (see e.g. \cite{seljak2006,McDonald:2004eu,Arinyo-i-Prats:2015vqa}).
We have thus considered five different values for both $\sigma_8$ (in the interval $[0.754, 0.904]$) and $n_{\rm eff}$ (in the range $[-2.3474,-2.2674]$). 

Concerning the astrophysical parameters, we have varied the thermal history of the Intergalactic Medium (IGM) in the form of the
amplitude $T_0$ and the slope $\widetilde{\gamma}$ of its temperature-density
relation, generally parametrised as $T=T_0(1+\delta_\mathrm{IGM})^{\widetilde{\gamma}-1}$, with $\delta_{\mathrm{IGM}}$ being the IGM overdensity~\cite{hui97}.
Specifically, we have considered a set of three different temperatures at mean density, $T_0(z = 4.2) = 6000, 9200, 12600$~K, which
evolve with redshift, as well as a set of three values for the slope of the temperature-density relation, $\widetilde{\gamma}(z = 4.2) = 0.88, 1.24, 1.47$. The reference thermal history has been chosen to be defined by $T_0(z = 4.2) = 9200$ and $\widetilde{\gamma}(z = 4.2) = 1.47$, and it provides a good fit to observations, as demonstrated in Ref.~\cite{bolton17} where several hydrodynamical simulations with the same reference thermal history as the one used here have been carried on.

We have also varied the redshift of the instantaneous reionization
model, for which we have considered the three different values $z_{\rm reio} = 7,9,15$, with $z_{\rm reio} = 9$ being the reference value. Furthermore, we have considered ultraviolet (UV) fluctuations of the ionizing background, that may be particularly important at high redshift. The amplitude of this effect is described by the parameter $f_{\rm UV}$: the corresponding template is built from a set of 3 models with $f_{\rm UV} = 0, 0.5, 1$, where
$f_{\rm UV} = 0$ corresponds to a spatially uniform UV background~\cite{Irsic:2017ixq}.

Finally, we have varied the mean flux $\bar{F}(z)$ by
selecting 9 different values for it, namely $(0.6,0.7,0.8,0.9,1.0,1.1,1.2,1.3,1.4) \times \bar{F}_{\rm REF}$, with the reference values being the ones of the SDSS-III/BOSS measurements~\cite{boss2013}.  
Aiming to have a very fine grid in terms of mean fluxes, we have also included 8 additional values, obtained by rescaling the optical depth $\tau = -\ln\bar{F}$,~i.e.~$(0.6,0.7,0.8,0.9,1.1,1.2,1.3,1.4) \times \tau_{\rm REF}$.

In order to put constraints on the nCDM models, we have generated the initial conditions corresponding to the 109 $\{\alpha,\beta,\gamma\}$-combinations listed in Table~\ref{tab:params} by using a modified version of the numerical code {\tt 2LPTic}~\cite{Crocce:2006ve}, following the same approach adopted in Ref.~\cite{Murgia:2017lwo}. We have used these snapshots as inputs for running 109 full hydrodynamical simulations (512$^3$ particles in a 20 Mpc$/h$ box), keeping the astrophysical and cosmological parameters fixed to their reference values.

Before investigating the fully general $\{\alpha,\beta,\gamma\}$-space, we wanted to be able to reproduce the same results obtained in Ref.~\cite{Irsic:2017ixq}, when the analysis is limited to the thermal WDM case. In order to do that, we have extended our grid of 109 nCDM points with 8 additional hydrodynamical simulations, in which the values for $\alpha$ correspond to thermal WDM masses of 2,3,4,5,6,7,8,9~keV, $\beta$ and $\gamma$ are fixed to their thermal values, and all the other cosmological and astrophysical parameters are fixed to their reference values. The full nCDM grid, including both thermal and non-thermal simulations, consists thereby in 117 points sampling the $\{\alpha,\beta,\gamma\}$-space.

\section{\label{sec:data}Data set}
In order to provide limits on the properties of nCDM, we have used a high resolution data set, constituted by the HIRES/MIKE samples of QSO spectra.
It has been obtained  with the HIRES/KECK and the MIKE/Magellan spectrographs, at redshift bins
$z=4.2,4.6,5.0,5.4$ and in 10 $k$-bins in the interval 0.001-0.08~s/km, with spectral resolution of 13.6 and 6.7 km/s, for HIRES and MIKE, respectively~\cite{Viel:2013apy}. As in the analyses of Refs.~\cite{Viel:2013apy,Irsic:2017ixq}, we have imposed a conservative cut on the flux power spectra obtained from MIKE/HIRES data, and only the measurements with $k > 0.005$ s/km have been used, in order to avoid possible systematic uncertainties on large scales due to continuum fitting. Furthermore, we do not consider the highest redshift bin for MIKE data, for which the error bars on the flux power spectra are very large (see Ref.~\cite{Viel:2013apy} for more details).
We have used a total of of 49 $(k,z)$ data points for the MIKE/HIRES data set, which has the advantage with respect to other surveys of exploring small scales and high redshifts, being thereby the most constraining up-to-date, for the models that we aim to test. 

Note that low resolution surveys such as SDSS-III/BOSS~\cite{boss2013} can be used for constraining nCDM scenarios in the quasi-linear regime which characterises larger scales with respect to the ones that we have studied in this work. An interesting attempt of modelling the relevant features of the flux power spectrum in order to obtain an approximate estimator for testing such relatively large scales has been recently done by the authors of Ref.~\cite{Garny:2018byk}. Let us stress that the two approaches are complementary, since the different scale and redshift coverage may lead to different constraints and degeneracies, with the common goal of developing an effective framework which does not require to run specific numerical simulations per each nCDM model.

\section{\label{sec:method} Method}
With the models of the flux power spectra obtained from the large suite of hydrodynamical
simulations presented in Section~\ref{sec:sims}, we have set a sparse grid of points in the multidimensional parameter space of $\{\bar{F}(z), T_0(z), \widetilde{\gamma}(z), \sigma_8, z_{\rm reio}, n_{\rm eff}, f_{\rm UV}, \alpha, \beta, \gamma \}$.
For the interpolation between different grid points, we have adopted an improved method with respect to the linear interpolation scheme used in Refs.~\cite{Irsic:2017ixq,Irsic:2017yje}, i.e.~the \emph{Ordinary Kriging} method, which is widely used in very different fields from cosmology, such as geo-statistics or environmental science, since it is particularly effective for dealing with sparse and non-regular grids (see, e.g.,~\cite{webster2007geostatistics}).
The interpolation is done in terms of ratios between the flux power spectra of the nCDM models and the reference one. We first interpolate in the astrophysical and cosmological parameter space for the $\Lambda$CDM case,~i.e.~in the $\alpha=0$ plane. We then correct all the $\{\alpha, \beta, \gamma \}$-grid points accordingly, and we finally interpolate in the $\{\alpha, \beta, \gamma \}$-space. This procedure relies on the assumption that the corrections due to non-reference astrophysical or cosmological parameters are universal,~i.e.~we can apply the same corrections computed for the $\Lambda$CDM case ($\alpha=0$) to all the nCDM models described by our parametrisation.

We have carefully tested the new interpolation scheme, by iteratively predicting the value of the flux power spectrum at a given grid point without using that grid point, as well as by reproducing the bounds on the thermal WDM mass obtained in Ref.~\cite{Irsic:2017ixq}. 
Let us note that, in doing the latter, rather than using the full nCDM grid of 117 points, we have only used the 8 thermal WDM simulations, and hence applied the interpolation procedure to the same parameter space investigated in the previous analyses~\cite{Viel2005,Viel:2013apy,Irsic:2017ixq} (see Appendix~\ref{app:thermal}).
We have also run some additional simulations in order to furtherly test the accuracy of both the $\{\alpha, \beta, \gamma \}$-fitting procedure and the new interpolation scheme. The corresponding discussion is reported in Appendix~\ref{app:validation}.

Another difference with respect to the previous analyses is that we have not used cross simulations between the nCDM parameters and the astrophysical and cosmological ones. Notice, however, that the expected degeneracies,~e.g., between the IGM temperature and $\alpha$, have emerged consistently with respect to the results published so far (see Appendix~\ref{app:thermal}). However, we leave for future work a further extension of our parameter grid, by introducing such cross simulations.
On the other hand, we have noticed that the current interpolation scheme is not fully accurate for reproducing power spectra which are very far from the reference cases. This issue does not affect our final results on the nCDM parameters, having the only consequence of a further weakening of the bounds for those cosmological parameters which were not tightly constrained even in the previous studies (see Appendix~\ref{app:bf}).

In light of the aforementioned caveats, we have determined the constraints on both astrophysical, cosmological, and nCDM parameters, by maximising a Gaussian likelihood with a Monte Carlo Markov Chain (MCMC) approach, using the publicly available affine-invariant MCMC sampler \texttt{emcee}~\cite{emcee13}. 

Regarding the IGM thermal history, we have carried out two different analyses: following the approach of Ref.~\cite{Irsic:2017ixq}, we have modelled the redshift evolution of both $T_0$ and $\widetilde{\gamma}$ as power laws, such that $T_0(z) = T_0^A[(1+z)/(1+z_p)]^{T_0^S}$ and $\widetilde{\gamma}(z) = \widetilde{\gamma}^A[(1+z)/(1+z_p)]^{\widetilde{\gamma}^S}$, where the pivot redshift $z_p$ is the redshift at which most of the Lyman-$\alpha$ forest pixels are coming from (i.e.~$z_p = 4.5$ for MIKE/HIRES). We refer to this double power law parametrisation as our reference MCMC analysis.

However, in order to be conservative, we have repeated the same analysis by letting the amplitude $T_0(z)$ free to vary in each bin, only requiring to forbid differences greater than 5000~K between adjacent redshift bins~\cite{Viel:2013apy}. Furthermore, in order to prevent unreasonably cold values for the IGM temperatures, which would hardly be physically motivated, we have adopted broad gaussian priors centred on $T_0(z)$ reference values, with standard deviation $\sigma = 3000$~K.
As it has been thoroughly discussed in Ref.~\cite{Irsic:2017ixq}, different choices of the thermal history priors sensibly affect the results, due to the degeneracy between the IGM temperature evolution and the nCDM parameters (see also Ref.~\cite{Garzilli:2015iwa} for an analysis on the impact of very different thermal histories). When the power law evolution for $T_0$ is not assumed, the constraints on the small-scale power suppression associated to nCDM are expected to be weaker. For these reasons, in Appendix~\ref{app:thermal} we compare different prior choices on the IGM thermal history, both in the standard and the new framework, showing and discussing how they do influence the final results.

For the mean fluxes $\bar{F}(z)$ we have chosen gaussian priors with standard deviation $\sigma = 0.04$, approximately corresponding to the normalisation uncertainties given by different observations.
We adopted flat priors for all the other free parameters.

\section{\label{sec:results}Results and Discussion}

\begin{figure*}
\includegraphics[width=0.75\textwidth]{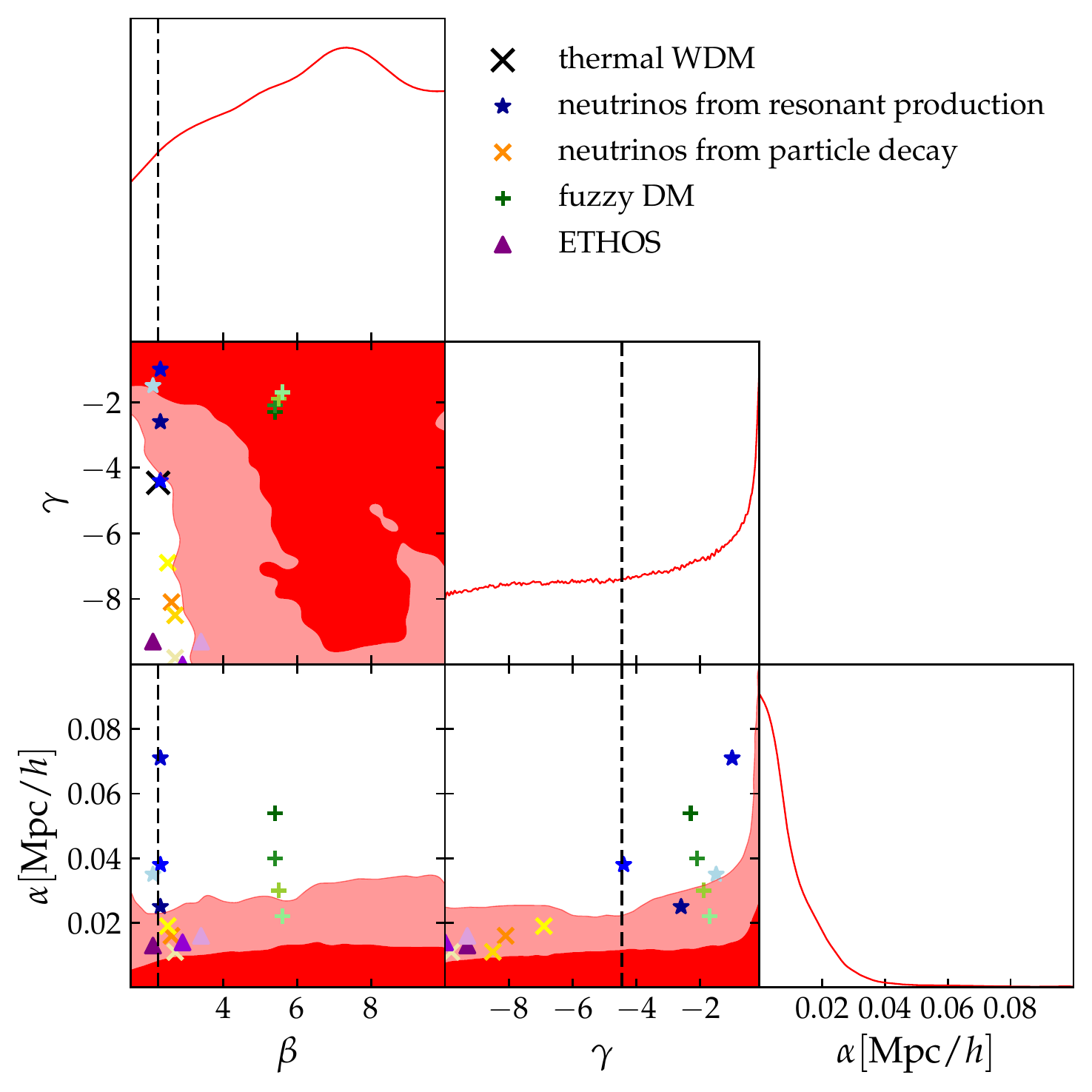}
\caption{Here we plot the 1$\sigma$ and 2$\sigma$ contour plots for $\alpha$, $\beta$ and $\gamma$, obtained by assuming a IGM temperature power law evolution (see the text for more details). The dashed vertical lines and the black cross stand for the thermal WDM case,~i.e.~$\beta=2.24$ and $\gamma=-4.46$. The other symbols shown in the legend correspond to the $\{\alpha, \beta, \gamma\}$-combinations associated to the nCDM models discussed in Ref.~\cite{Murgia:2017lwo} and listed in Table~\ref{tab:models}. Different colour gradients are used for distinguishing between different models belonging to the same class of nCDM scenarios. For each class, the darkest tonality corresponds to the first model listed in Table~\ref{tab:models}, the lightest one corresponds to the last model evaluated.}
\label{fig:ABG}
\end{figure*}

\begin{figure*}
\includegraphics[width=0.9\textwidth]{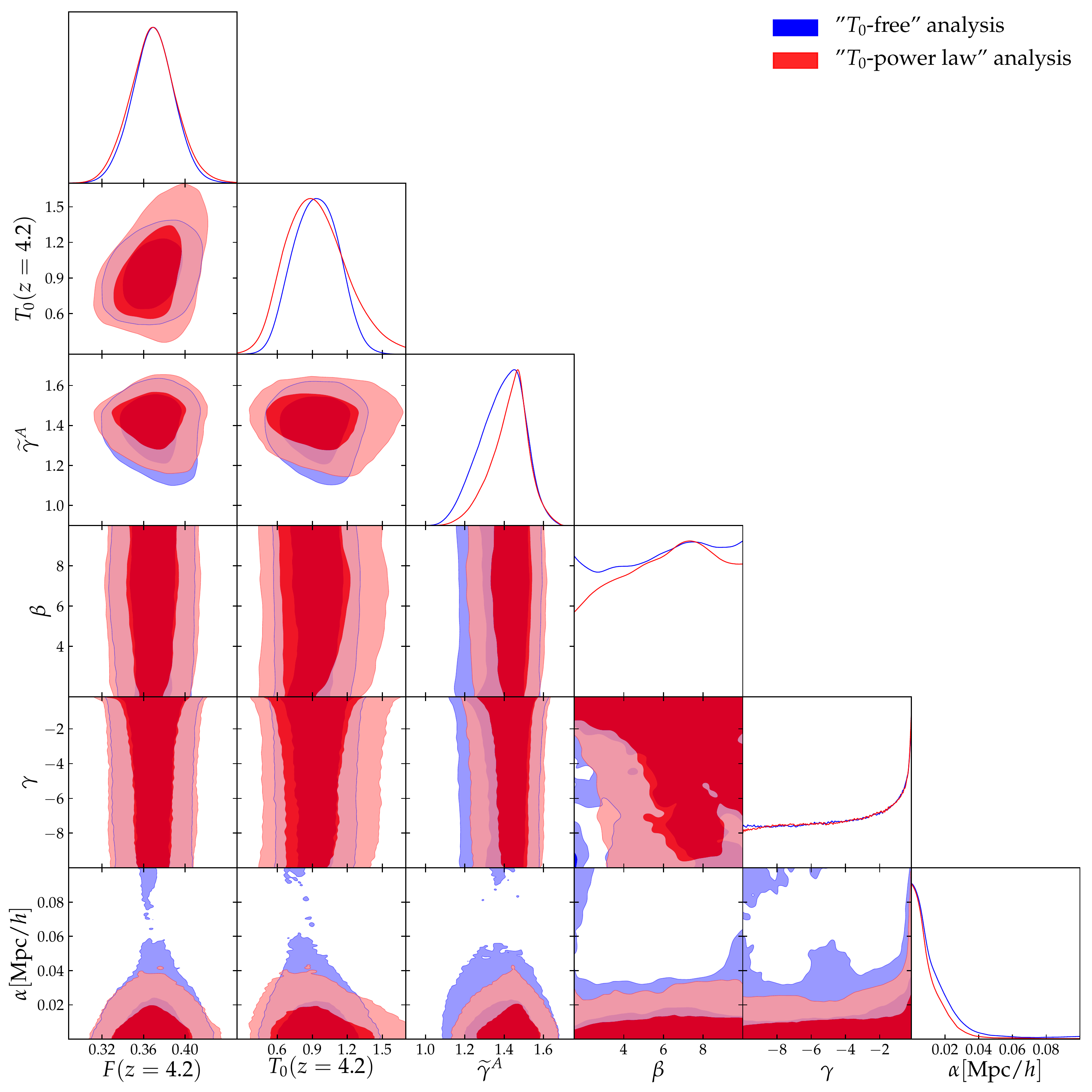}
\caption{Here we plot the 1$\sigma$ and 2$\sigma$ exclusion plots for both $\alpha$, $\beta$, $\gamma$, and the main astrophysical free parameters. The blue contours refer to the freely floating IGM temperature analysis, while the red contours refer to our reference analysis,~i.e.~when a power law evolution is assumed. The values of the temperature are expressed in $10^{4}$~K units.}
\label{fig:ABG2}
\end{figure*}

\begin{figure}
\includegraphics[width=0.48\textwidth]{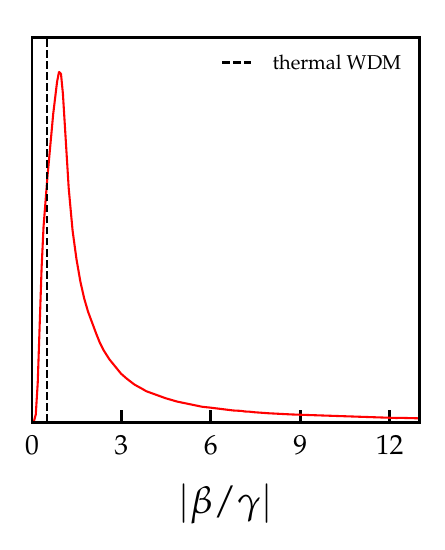}
\caption{Here we plot the marginalised 1D distribution of $|\beta / \gamma|$, which is a useful estimator for condensing the constraints on the two parameters governing the slope of the power suppression. The vertical dashed line corresponds to the thermal WDM $\{\beta,\gamma\}$-combination,~i.e.~$\beta=2.24$ and $\gamma=-4.46$.} 
\label{fig:bg}
\end{figure}

In this Section we discuss the results of the comprehensive MCMC analyses that we have performed for the MIKE/HIRES data set. Firstly, we have focused on the thermal WDM case, by switching off the parameters responsible of non-trivial features in the shape of the small-scale power cut-off (i.e.$~\beta$ and $\gamma$), and thus by constraining the same parameter space studied in Ref.~\cite{Viel:2013apy,Irsic:2017ixq}, where $\alpha$ plays the role of the thermal WDM mass parameter. In Appendix~\ref{app:thermal} we compare our results for this specific class of nCDM models with the constraints previously published, showing the full consistency between them.

In this Section, let us then focus on the main goal of this work,~i.e.~putting limits on the $\{\alpha, \beta, \gamma\}$-space. In Fig.~\ref{fig:ABG} we condense our main results, namely the 1$\sigma$ and 2$\sigma$ exclusion plots showing the bounds on the three parameters describing the power suppression induced by nCDM. We have chosen to focus on the analysis based on the assumption of a power law evolution for both the amplitude and the slope of the IGM temperature, which is also the case adopted as reference in Ref.~\cite{Irsic:2017ixq}, due to its robust physical motivations. However, as it is manifestly shown in Fig.~\ref{fig:ABG2}, the more conservative assumption of a thermal history with freely floating $T_0(z)$ does not change the conclusions discussed here.

From Fig.~\ref{fig:ABG} we note that, even in our new general framework, the parameter $\alpha$, responsible of the position of the cut-off in the power spectrum is well constrained by current data. On the other hand, both $\beta$ and $\gamma$ are quite unconstrained. It is interesting to notice, however, that the 1D posterior distribution of the former shows a peak around $\beta=7$, which is far from its thermal value,~i.e.~$\beta=2.24$. The natural interpretation is that standard thermal WDM models are not favoured by Lyman-$\alpha$ data with respect to non-thermal scenarios. In the plots, the thermal values for $\beta$ and $\gamma$ are highlighted by the dashed vertical lines and the black cross. More specifically, the black cross, which corresponds thereby to $\beta=2.24$ and $\gamma=-4.46$, lies slightly outside of the 2$\sigma$ contour. The peak in the 1D distribution of $\alpha$ corresponds to the standard CDM model, and the mild degeneracy between large values of $\alpha$ and small values of $|\gamma|$ was predicted and extensively discussed in our previous work~\cite{Murgia:2017lwo}.

In Ref.~\cite{Murgia:2017lwo} we also discussed several viable classes of nCDM models motivated by particle physics (i.e., sterile neutrinos both from resonant production and particle decay, fuzzy DM models, mixed DM fluids, DM-Dark Radiation interaction models), and we analysed some examples from each of the families. The symbols reported in Fig.~\ref{fig:ABG} correspond to the $\{\alpha, \beta, \gamma\}$-combinations which have been shown to provide a good fit for the transfer functions associated to such example models.
In order to quantify their viability, we list all of them, with the corresponding $\alpha, \beta, \gamma$ and $k_{1/2}$, in Table~\ref{tab:models}, where we report the $\chi^2$ values determined through our reference analysis. Clearly, the $\chi^2$ values have been computed only for those models associated to $\{\alpha, \beta, \gamma\}$-combinations sampling a parameter region which is covered by our grid of simulations (see Table~\ref{tab:params}).
In Fig.~\ref{fig:ABG}, different colour gradients are used for distinguishing between different models belonging to the same class of nCDM scenarios. For each group of models, the darkest tonality corresponds to the first model listed in Table~\ref{tab:models}, while the lightest one corresponds to the last model evaluated.
 
We address the reader to Ref.~\cite{Murgia:2017lwo} for a more detailed treatment of the specific physical features of the different particle physics scenarios listed in Table~\ref{tab:models}. However, let us shortly recall that the first group of models corresponds to few example values for the resonantly produced sterile neutrino mass ($m_{\rm{S}}=5,7,15$~keV), the active-sterile neutrino mixing angle, and the lepton asymmetry. Concerning sterile neutrinos from particle decay, each of the model is characterised by different values of the sterile neutrino and decaying scalar particle masses, along with different Higgs portal and Yukawa coupling parameters. Fuzzy DM scenarios rely on the assumption that all the DM is constituted by an ultra-light scalar particle: the first two models belonging to this class correspond to DM masses of $5 \cdot 10^{-22}$ and $10 \cdot 10^{-22}$~eV, and they are rejected by our analysis; the latter two fuzzy DM scenarios correspond to masses $20 \cdot 10^{-22}$ and $40 \cdot 10^{-22}$~eV, and they are accepted.

Let us stress that, thanks to the new general parametrisation, it has not been necessary to run any specific numerical simulations in order to test such nCDM scenarios. Whenever one wants to constrain any model belonging to one of these families, it is sufficient to fit the corresponding linear power spectrum in terms of $\{\alpha, \beta, \gamma\}$, and interpolate in the parameter space delineated by our full grid of simulations,~i.e.~in a refined $\chi^2$-table having a similar structure to Table~\ref{tab:models}, but also including all the astrophysical and cosmological parameters involved.

By looking at the positions of the various symbols in Fig.~\ref{fig:ABG}, one cannot give a definitive answer about the viability of the whole classes of candidates.
This aspect is particularly relevant for the models suggested by the effective theory of structure formation (ETHOS), which often feature oscillations at very small scales, that our parametrisation cannot capture. For the few examples considered here, all referring to a weak Dark Acoustic Oscillation (DAO) regime, the presence of such oscillations is totally negligible for the data analyses, as it is explicitly displayed in Appendix~\ref{app:validation}. We leave for a future work a deeper investigation for quantifying to which extent our fitting procedure is able to cover the whole class of ETHOS models, both in the weak and in the strong DAO regime (see Refs.~\cite{Cyr-Racine:2015ihg,Vogelsberger:2015gpr} for further details). A similar issue concerns fuzzy DM models, which are expected to modify the dynamics during the non-linear phase of structure formation, due to quantum pressure effects. Concerning this point, we address the reader to Ref.~\cite{Nori:2018pka}, where it has been probed for the first time that such effects do not affect predictions obtained under the standard approximation of treating ultra-light scalars as standard collisionless DM, at least for models where fuzzy DM constitutes the whole DM amount. 

Taking into consideration all the points above, it is now possible to see which nCDM models are excluded by current data and which ones are not.
The fuzzy DM model examples that we have considered (green crosses) are associated to values for $\beta$ and $\gamma$ which are in perfect agreement with data. However, from Fig.~\ref{fig:ABG} it is manifest that only the values of $\alpha$ relative to the last two models reported in Table~\ref{tab:models},~i.e.~those featuring a power suppression at the smallest scales, are allowed by our analysis. This means that a DM fluid fully composed by an ultra-light scalar field is characterised by a shape for the power suppression which can always accommodate Lyman-$\alpha$ data, provided that the mass of the scalar particle is sufficiently large. Analogous arguments apply to the four resonantly produced sterile neutrino models that we have tested (blue stars).
Conversely, by looking at the positions of the orange and yellow crosses in Fig.~\ref{fig:ABG} we can conclude that, even though the power suppression due to sterile neutrinos from particle decay occurs at scales allowed by data, the corresponding $\{\beta,\gamma\}$-combinations lie at the border of the 2$\sigma$ contour, displaying that it is the peculiar shape of the power suppression to slightly disfavour this class of models.
Concerning the ETHOS example models that we have analysed (purple triangles), it is interesting to note that, whereas all the corresponding values of $\alpha$ are in agreement with data, the $\{\beta,\gamma\}$-combinations associated to the first two scenarios are rejected. Such combinations lead indeed to a power suppression at relatively large scales, as it is quantified by the corresponding $k_{1/2}$ values listed in Table~\ref{tab:models}.

To summarise our main message, thanks to the new parametrisation it is now possible to test a wide variety of nCDM models with Lyman-$\alpha$ data, by simply fitting their linear power spectra in terms of $\{\alpha, \beta, \gamma\}$, and confronting the corresponding $\{\alpha, \beta, \gamma\}$-combinations against exclusion plots like the ones shown in Fig.~\ref{fig:ABG}. In other words, it is sufficient to look at which region of the parameter space they sample, without the need of running any numerical simulations.

Now we would like to stress that none of the mixed (cold + warm) DM scenarios discussed in our previous work is overplotted in Fig.~\ref{fig:ABG}. This is due to the fact the values of $\alpha$ which are needed in order to fit the transfer functions associated to such models must necessarily be greater than 0.1, well beyond the constraint that we have obtained.
By marginalising over all the other parameters, we obtain indeed an upper limit on $\alpha < 0.03~{\rm{Mpc}}/h$ (2$\sigma$~C.L.), which could be interpreted as the largest possible scale at which a power suppression induced by any nCDM scenario can be present, in order to be in agreement with Lyman-$\alpha$ data, provided that such nCDM scenario is captured by our parametrisation.
Such constraint constitutes a strong hint that mixed DM fluids composed by a cold plus a warm (thermal) component are disfavoured by structure formation data. A more comprehensive and systematic study focused on this particular class of models is needed before claiming that they are completely ruled out. Nevertheless, in light of our analyses it is clear that only scenarios with large masses and/or tiny abundances for the warm (thermal) component may accommodate Lyman-$\alpha$ data,~i.e.~scenarios which are practically indistinguishable with respect to the standard CDM model, with current structure formation observations.

In Fig.~\ref{fig:ABG2} we plot the 1D and 2D distributions for both $\alpha$, $\beta$, $\gamma$, and the main astrophysical free parameters. The blue contours refer to the freely floating IGM temperature analysis, while the red contours refer to the case where a power law evolution is assumed.
Both in Section~\ref{sec:method} and in Appendix~\ref{app:thermal} we extensively discuss how different prior choices on the IGM temperature evolution influence the constraints on the WDM mass, when the analysis is limited to thermal models. Interestingly, the consequences of such different choices on $\alpha$ are mitigated in the more general $\{\alpha, \beta, \gamma\}$-analysis, where the effects of different assumptions on the thermal history are somehow spread on the distributions of the three parameters associated to the nCDM nature. By examining the contour plots shown in Fig.~\ref{fig:thermal_Tpowlaw}, the relative stability of the 1D distribution of $\alpha$ is visibly evident. However, by dropping the assumption of a temperature power law evolution and letting $T_0(z)$ free to vary bin by bin, the marginalised 2$\sigma$ limit on $\alpha$ is weakened, being $\alpha < 0.05~{\rm{Mpc}}/h$.

\begin{table}
\setlength{\tabcolsep}{3pt}\renewcommand{\arraystretch}{1.2}
\begin{tabular}{|c||c|c|c|c||c|} \hline
& {$\alpha~[{\rm Mpc}/h]$} &  {$\beta$} & {$\gamma$} & {$k_{1/2}~[h/{\rm Mpc}]$} & {$\chi^2$} \\ \hline\hline
\scriptsize{\bf Neutrinos} & {$0.025$} & {$2.3$} & {$-2.6$} & {17.276}& 101\\
\scriptsize{\bf from} & {$0.071$} & {$2.3$} & {$-1.0$} & {9.828} & 266\\
\scriptsize{\bf resonant} & {$0.038$} & {$2.3$} & {$-4.4$} & {8.604} & 283\\
\scriptsize{\bf production} & {$0.035$} & {$2.1$} & {$-1.5$} & {15.073} &  149\\
\hline                                                                                                                                                                         
\scriptsize{\bf Neutrinos} & {$0.016$} & {$2.6$} & {$-8.1$} & {19.012} & 104\\
\scriptsize{\bf from} & {\bf0.011} & {\bf2.7} & {\bf-8.5} & {\bf28.647} & \bf38\\
\scriptsize{\bf particle} & {$0.019$} & {$2.5$} & {$-6.9$} & {16.478} &  105\\
\scriptsize{\bf decay}	& {\bf0.011} & {\bf2.7} & {\bf-9.8} & {\bf26.31} &  \bf45\\
\hline                                                                                                                                                                                                
\scriptsize{} & {$0.16$} & {$3.2$} & {$-0.4$} & {6.743} &  229\\
\scriptsize{\bf Mixed} & {$0.20$} & {$3.7$} & {$-0.18$} & {7.931} &  -\\
\scriptsize{\bf models} & {$0.21$} & {$3.7$} & {$-0.1$} & {11.36} &  -\\
\scriptsize{} & {$0.21$} & {$3.4$} & {$-0.053$} & {33.251} &  -\\
\hline                                                                                                                                                            
\scriptsize{} & {$0.054$} & {$5.4$} & {$-2.3$} & {13.116} & 169\\
\scriptsize{\bf Fuzzy} &  {$0.040$} & {$5.4$} & {$-2.1$} & {18.106} & 104\\
\scriptsize{\bf DM} & {\bf0.030} & {\bf5.5} & {\bf-1.9} & {\bf25.016} & \bf40\\
\scriptsize{} & {\bf0.022} & {\bf5.6} & {\bf-1.7} & {\bf34.590} & \bf30\\
\hline                                                                                                                                                                            
\scriptsize{} & {$0.0072$} & {$1.1$} & {$-9.9$} & {7.274} & - \\
\scriptsize{\bf ETHOS} & {$0.013$} & {$2.1$} & {$-9.3$} & {16.880} & 153\\
\scriptsize{\bf models} & {$0.014$} & {$2.9$} & {$-10.0$} & {21.584} & 70\\
\scriptsize{} & {$0.016$} & {$3.4$} & {$-9.3$} & {23.045} & 60\\ \hline
\end{tabular}
\caption{Here we list 20 $\{\alpha, \beta, \gamma\}$-combinations, with the corresponding value for $k_{1/2}$, each of them referring to one of the nCDM particle models examined in Ref.~\cite{Murgia:2017lwo} and overplotted in Fig.~\ref{fig:ABG}. They are split into five groups, which represent some of the most viable classes of nCDM scenarios up-to-date. In the last column we report the corresponding $\chi^2$ values from our reference analysis. Those models for which the $\chi^2$ is not shown are associated with $\{\alpha, \beta, \gamma\}$-combinations sampling a parameter region not covered by our grid of simulations (see Table~\ref{tab:params}). Models highlighted in bold-face are accepted (at 2$\sigma$~C.L.) by our analysis.
Notice that, thanks to the new parametrisation, in order to test against Lyman-$\alpha$ data any model belonging to one of these groups, it is sufficient to fit its linear power spectrum in terms of $\{\alpha, \beta, \gamma\}$, and interpolate in a similar $\chi^2$-table, without the need to run any numerical simulation (see the text for details).\label{tab:models}}
\end{table}

It is also informative to look at the marginalised 1D distribution of the quantity $|\beta / \gamma|$, plotted in Fig.~\ref{fig:bg}, which somehow compresses the information about the slope of the power suppression, and appears to be well constrained by our analysis. The corresponding $2\sigma$ upper limit, obtained by marginalising over $\alpha$, is the following: $|\beta / \gamma| < 14$. The vertical dashed line refers to the position of the $\{\beta, \gamma\}$-combination associated to thermal WDM models. Consistently with what we have already pointed out, its position shows that the particular shape of the power suppression induced by a thermal WDM candidate is not favoured by data. 

As it was expected, the bounds on the astrophysical and cosmological parameters do not present dramatic differences with respect to the thermal WDM case, apart from a mild overall weakening due to the addition of two more free parameters to the analysis. The effects on the flux power spectra induced by variations of most of the astrophysical and cosmological parameters, in fact, mainly occur at larger scales with respect to the ones influenced by the nCDM parameters, with the remarkable exception,~e.g., of the IGM temperature. 

Lastly, in Fig.~\ref{fig:Tk2s} we explicitly show, among the 109 $\{\alpha, \beta, \gamma\}$-combinations listed in Table~\ref{tab:params}, the squared transfer functions corresponding to nCDM models which are accepted at $2\sigma$ C.L. by our reference analysis (red lines) and the ones associated to models that are rejected (blue lines). The nCDM models corresponding to the red curves are associated to the $\{\alpha, \beta, \gamma\}$-triplets highlighted in bold-face in Table~\ref{tab:params}.

\begin{figure}
\includegraphics[width=0.49\textwidth]{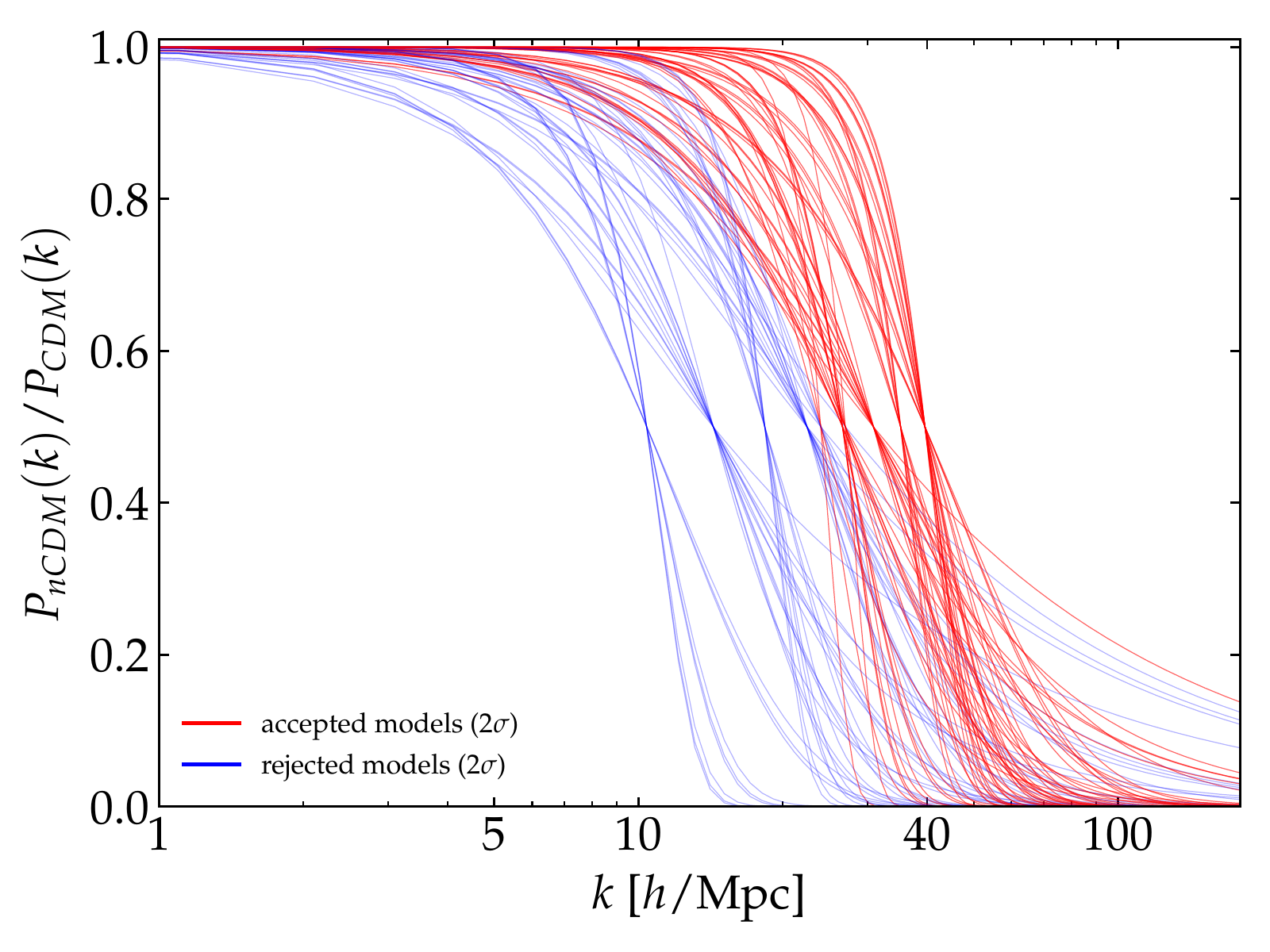}
\caption{Here we plot the squared transfer functions associated to the 109 $\{\alpha,\beta,\gamma\}$-combinations listed in Table~\ref{tab:params}, and we highlight which ones are accepted by our reference MCMC analysis at $2\sigma$ C.L. (red lines) and which ones are rejected (blue lines).}
\label{fig:Tk2s}
\end{figure}

\section{\label{sec:conclusions}Conclusions}
In the past, considerable efforts have gone into studying the observational implications of a very specific class of DM models, the so-called thermal WDM models,~i.e.~candidates featuring a thermal momentum distribution. In the standard framework, the imprint on the matter power spectrum due to the presence of nCDM is parametrised through a one-to-one correspondence between the position of the small-scale suppression with respect to the standard CDM model, and the mass of the thermal WDM candidate. However, the suppression in the matter power spectrum induced by nCDM can feature different strength and shape, based on the fundamental nature of the DM candidate. Since most of the viable nCDM scenarios provided by theoretical particle physics are not adequately described by the oversimplified thermal WDM case, in Ref.~\cite{Murgia:2017lwo} we introduced a more general parametrisation, which allows to reproduce the detailed features of the power suppression produced by the most viable non-thermal DM scenarios. Well motivated nCDM particle models such as sterile neutrinos, fuzzy DM, DM-Dark radiation interaction models, mixed DM fluids, are efficiently covered by the novel parametrisation, thanks to the mutual dependence among its three free parameters $\alpha$, $\beta$ and $\gamma$.

In this work we have presented the first accurate constraints on the three parameters characterising the new framework, provided by an extensive MCMC analysis of the high resolution and high redshift MIKE/HIRES Lyman-$\alpha$ forest data. Our results rely on a large set of full hydrodynamical simulations, which constitutes a robust template of mock flux power spectra to be confronted against observations.

Firstly, we have shown that when our analysis is limited to the thermal WDM case, we recover the results currently available in the literature.
We have then extracted absolute limits on the $\{\alpha,\beta,\gamma\}$-space, determining a marginalised upper limit on the largest possible scale at which a significant power suppression induced by any nCDM scenario can occur,~i.e.~$\alpha<0.03~{\rm{Mpc}}/h$ (2$\sigma$~C.L.), provided that such scenario can be fitted in terms of $\{\alpha,\beta,\gamma\}$.

We have also examined several specific examples among the aforementioned non-thermal nCDM models, and we have illustrated how to test them against Lyman-$\alpha$ data, without the need to perform any dedicated numerical simulations. We have shown that, thanks to the versatility of the new parametrisation, in order to constrain the fundamental properties of a nCDM scenario, it is sufficient to fit its linear power spectrum in terms of $\{\alpha,\beta,\gamma\}$, and interpolate in the grid of flux power spectra presented in this work.

We have explicitly shown that standard thermal WDM models are not favoured by Lyman-$\alpha$ data with respect to non-thermal scenarios.
Even though a more systematic sampling of the parameter space is required for drawing definitive conclusions, we have also claimed that a DM fluid composed by both a warm (thermal) and a cold component is in tension with the Lyman-$\alpha$ forest, unless it is nearly indistinguishable from the standard CDM model, with current observational data.

Future developments and improvements of this project will firstly consist in extending and refining our multidimensional parameter grid, by adding hydrodynamical simulations in which,~e.g.,~the astrophysical and the nCDM parameters are varied simultaneously. Furthermore, it may be interesting to focus on refining specific regions of the $\{\alpha,\beta,\gamma\}$-grid, associated to particularly viable nCDM scenarios.
We are also working on expanding the data set for the analysis, by adding the sample of medium resolution and intermediate signal-to-noise QSO spectra obtained by the XQ-100 survey~\cite{xq100}. At it has been already stressed, in fact, combining data sets with complementary redshift and scale coverages is expected to break some degeneracies and possibly tighten the constraints presented here.

Since the present data, especially in the high redshift and small scale regime, rely on about 20 QSO spectra, it is expected that by increasing the size of the data set the present constraints can be improved. Moreover, an independent and accurate measurement of the thermal history would provide strong priors on the thermal parameters that in turn will allow to better constrain the shape of the linear power spectrum at such small scales.

Lastly, we are planning to make public a simple code, which will allow the user to fit with our parametrisation a given linear power spectrum, interpolate in the $\{\alpha,\beta,\gamma\}$-space, and automatically translate the limits on the three parameters to bounds on the corresponding fundamental nCDM model.

\appendix

\section{\label{app:grid} The grid of nCDM simulations}
\begin{table}
  \setlength{\tabcolsep}{5pt}
  \begin{tabular}{|c|c|c ||| c|c|c|} \hline
  \small{$\alpha$}~\scriptsize{$[{\rm Mpc}/h]$} & \small{$\beta$} & \small{$\gamma$} & \small{$\alpha$}~\scriptsize{$[{\rm Mpc}/h]$} & \small{$\beta$} & \small{$\gamma$} \\ \hline
0.008 & 1.5 & -10.00 & 0.023 &	 2.0 &	 -6.00  \\
0.005 &	 1.5 & -10.00 & \bf0.009 &	\bf2.0 & \bf-6.00  \\
\bf0.003 &	\bf1.5 & \bf-10.00 &\bf 0.006 &	\bf 2.0 &\bf-6.00  \\
0.012 &	 1.5 & -5.00  & 0.029 &	 2.0 &	 -4.00  \\
0.008 &	 1.5 & -5.00  & \bf0.011 &	\bf 2.0 & \bf-4.00  \\
\bf0.006 &	\bf1.5 & \bf-5.00 & \bf0.008 & \bf2.0 &	\bf-4.00  \\
0.039 &	 1.5 & -1.00  & 0.042 &	2.0 &	-2.00  \\
0.025 &	 1.5 & -1.00  & \bf0.016 &	 \bf2.0 &	 \bf-2.00  \\
\bf0.018 &	\bf1.5 & \bf-1.00  & \bf0.011 &	\bf2.0 &\bf-2.00  \\
0.013 &	 2.0 & -10.00 & 0.047 &	 4.0 &	 -6.00  \\
0.008 &	 2.0 & -10.00 & \bf0.019 &	\bf 4.0 &\bf-6.00  \\
\bf0.006 &	\bf2.0 & \bf-10.00 & \bf0.012 &\bf4.0 &	\bf-6.00  \\
0.019 &	 2.0 & -5.00  & 0.053 &	 4.0 &	 -4.00  \\
0.012 &	2.0 & -5.00  & \bf0.021 & \bf4.0 & \bf-4.00  \\
\bf0.009 &	\bf2.0 & \bf-5.00  & \bf0.014 &	\bf4.0 &\bf-4.00  \\
0.045 &	 2.0 & -1.00  & 0.063 &	 4.0 &	 -2.00  \\
0.029 &	2.0 & -1.00  & \bf0.025 &\bf4.0 &\bf-2.00  \\
\bf0.021 &	\bf2.0 & \bf-1.00  & \bf0.017 &	\bf4.0 &\bf-2.00  \\
0.018 &	 2.5 & -10.00 & 0.060 &	 6.0 &	 -6.00  \\
0.012 &	2.5 & -10.00 & \bf0.023 &\bf6.0 &\bf-6.00  \\
\bf0.008 &	\bf2.5 & \bf-10.00 & \bf0.016 &	 \bf6.0 &\bf-6.00 \\
0.024 &	 2.5 & -5.00  & 0.064 &	 6.0 &	 -4.00  \\
0.016 &	 2.5 & -5.00  & \bf0.025 &\bf6.0 &\bf-4.00  \\
\bf0.011 &	\bf2.5 & \bf-5.00  & \bf0.017 &	\bf6.0 &\bf-4.00  \\
0.049 &	 2.5 & -1.00  & 0.073 &	 6.0 &	 -2.00  \\
0.031 &	2.5 & -1.00  & \bf0.028 & \bf6.0 & \bf-2.00\\
\bf0.023 &	\bf2.5 & \bf-1.00  & \bf0.019 &	\bf 6.0 &\bf-2.00  \\
0.011 &	 2.0 & -5.00  & 0.020 &	 3.0 &	 -7.50  \\
\bf0.010 &	 \bf2.0 & \bf-5.00  & \bf0.010 &\bf3.0 &\bf-7.50  \\
0.015 &	 2.5 & -5.00  & \bf0.009 &\bf3.0 &\bf-7.50  \\
\bf0.013 &	 \bf2.5 & \bf-5.00  & 0.029 &	 3.0 &	 -2.50  \\
\bf0.025 &	 \bf5.0 & \bf-5.00  & \bf0.015 &\bf3.0 &\bf-2.50  \\
\bf0.022 &	 \bf5.0 & \bf-5.00  & \bf0.013 &\bf3.0 &\bf-2.50  \\
\bf0.032 &	 \bf10.0& \bf-5.00  & 0.041 &	 3.0 &	 -1.00  \\
\bf0.028 &	 \bf10.0& \bf-5.00  & \bf0.021 &\bf3.0 &\bf-1.00  \\
0.095 &	 2.5 & -0.30  & \bf0.019 &	\bf3.0 & \bf-1.00  \\
0.169 &	 2.5 & -0.15  & 0.030 &	 5.0 &	 -7.50  \\
0.061 &	 2.5 & -0.30  & \bf0.015 &\bf5.0 &\bf-7.50  \\
0.108 &	 2.5 & -0.15  & \bf0.014 &\bf5.0 &	 \bf-7.50  \\
\bf0.044 &	 \bf2.5 & \bf-0.30  & 0.037 &	 5.0 &	 -2.50  \\
\bf0.078 &	 \bf2.5 & \bf-0.15  & \bf0.019 &\bf5.0 &\bf-2.50  \\
0.057 &	 2.5 & -0.30  & \bf0.017 &\bf 5.0 &\bf-2.50  \\
0.101 &	2.5 & -0.15  & 0.046 & 5.0 &-1.00  \\
\bf0.051 &	\bf 2.5 & \bf-0.30  & \bf0.024 &\bf 5.0 &\bf-1.00  \\
\bf0.090 &	\bf 2.5 & \bf-0.15  & \bf0.021 &\bf 5.0 &\bf-1.00  \\
0.082 &	5.0 & -0.30  & 0.035 & 7.0 &-7.50  \\
0.109 &	5.0 & -0.15  & \bf0.018 &\bf 7.0 &\bf-7.50  \\
0.052 &	 5.0 & -0.30  & \bf0.016 &\bf7.0 &\bf-7.50  \\
0.069 &	5.0 & -0.15  & 0.042 &7.0 &-2.50  \\
\bf0.038 &	\bf5.0 & \bf-0.30  & \bf0.022 &	 \bf7.0 &\bf-2.50  \\
\bf0.050 &	 \bf5.0 & \bf-0.15  & \bf0.019 &\bf7.0 &	\bf-2.50  \\
\bf0.049 &	\bf 5.0 & \bf-0.30  & 0.048 &	 7.0 &	 -1.00  \\
0.065 &	 5.0 & -0.15  & \bf0.025 &\bf7.0 &\bf-1.00  \\
\bf0.043 &	\bf5.0 & \bf-0.30  & \bf0.022 &	\bf 7.0 &\bf-1.00  \\
\bf0.058 &	 \bf5.0 & \bf-0.15  &       &      &          \\ \hline
  \end{tabular}
 \caption{\label{tab:params}Here we report the 109 $\{\alpha,\beta,\gamma\}$-combinations that we have considered for our analyses, each of them associated to a different non-thermal nCDM model. We have used the corresponding transfer functions, computed via Eq.~\eqref{eq:Tgen}, as initial conditions for building our grid of hydrodynamical simulations. Models highlighted in bold-face are accepted at $2\sigma$~C.L. by our reference analysis.}
\end{table}

In Table~\ref{tab:params} we have listed the 109 $\{\alpha,\beta,\gamma\}$-combinations corresponding to the different non-thermal nCDM initial conditions that we have used as inputs for performing the hydrodynamical simulations constituting our grid in the $\{\alpha,\beta,\gamma\}$-space. As it is explained in Section~\ref{sec:sims}, the full nCDM grid includes also 8 additional points, each of them corresponding to a thermal WDM simulation, with WDM masses between 2 and 9 keV. Models highlighted in bold-face are accepted at $2\sigma$ C.L. by our reference analysis. They correspond indeed to the red transfer functions plotted in Fig.~\ref{fig:Tk2s}, and to the grey dots sampling the vertical shaded band shown in Fig.~\ref{fig:area}. Notice that large values for $\alpha$, corresponding to a power suppression happening at relatively large scales, are more likely to be allowed by data when $|\gamma|$ is very small,~i.e.~when the transfer function is shallower at the smallest scales.

\section{\label{app:thermal} Reproducing the thermal WDM limits}

\begin{figure}
\begin{minipage}{0.49\textwidth}
\vspace{0.2cm}
\subfloat[\label{fig:thermal_Tfree}]
{\includegraphics[width=\textwidth]{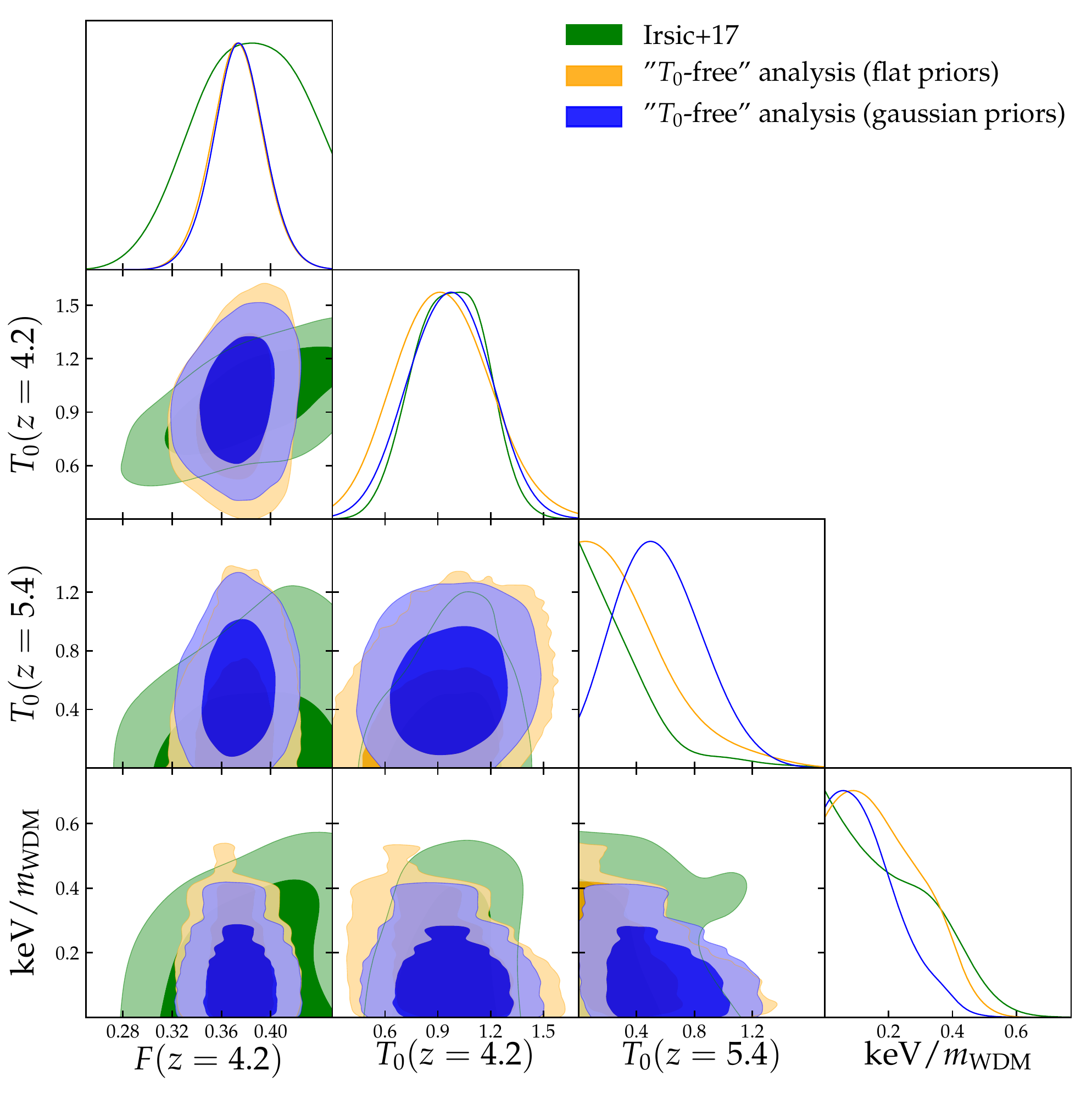}}
\vspace{0.5cm}
\end{minipage}
\begin{minipage}{0.49\textwidth}
\subfloat[\label{fig:thermal_Tpowlaw}]
{\includegraphics[width=\textwidth]{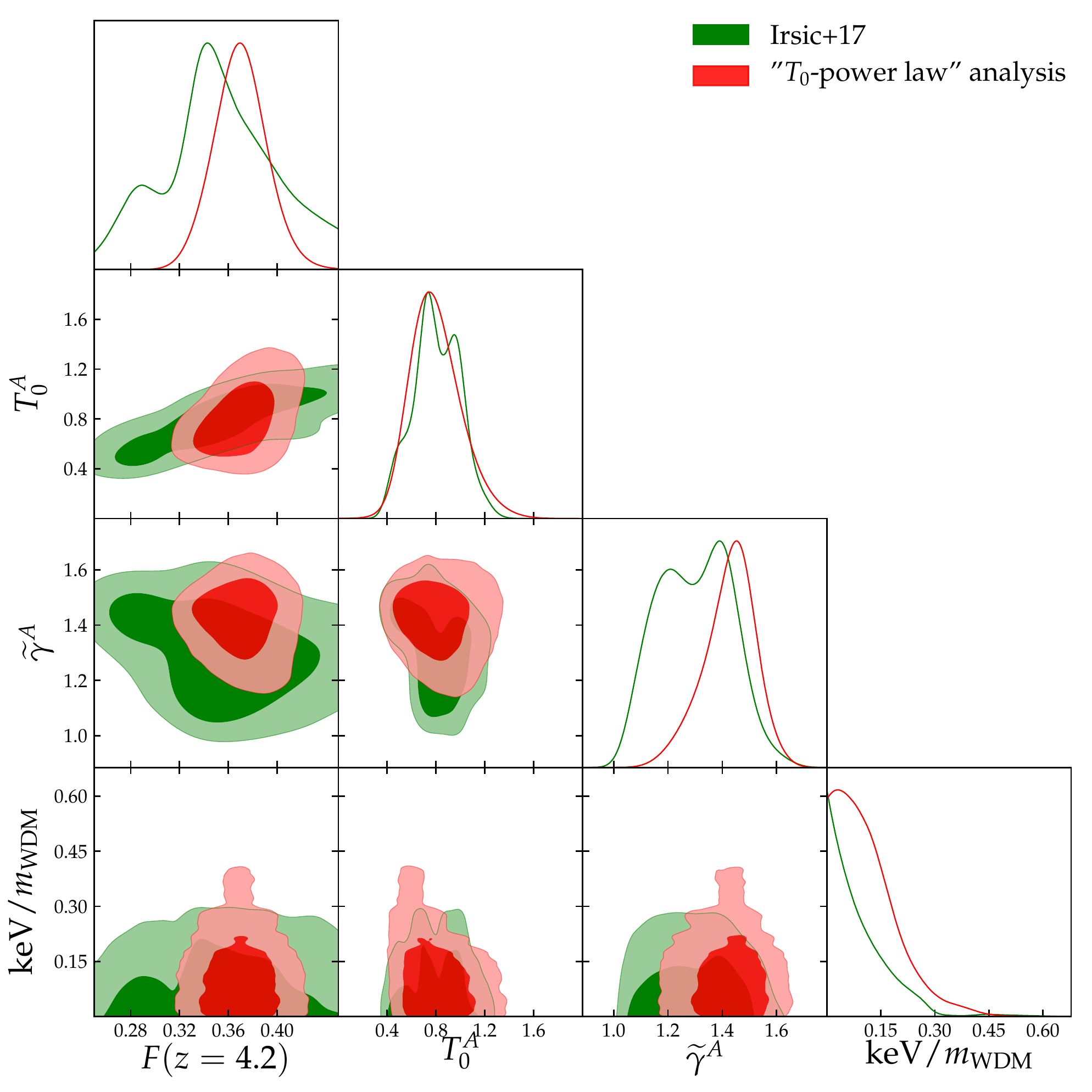}}
\vspace{0.1cm}
\end{minipage}
\caption{Here we compare the 1$\sigma$ and $2\sigma$ exclusion plots obtained with our thermal WDM analyses against the results from Ref.~\cite{Irsic:2017ixq}, which are displayed as green contours. The values of the temperatures are expressed in $10^{4}$~K units.~(a)~In the top panel we focus on the freely floating temperature analysis (blue and orange contours).~(b)~In the bottom panel we focus on the power law analysis (red contours).}
\end{figure}

This Appendix is dedicated to a comparison between the predictions that we have obtained, when limiting our analysis to the thermal WDM case, and the most updated published results obtained with the standard approach~\cite{Irsic:2017ixq}. The goal is to check the accuracy of the new interpolation scheme and sampling method, in order to safely extend our analyses to the full $\{\alpha, \beta, \gamma\}$-space.

In Fig.~\ref{fig:thermal_Tfree} we report a comparison between the 1D and 2D posterior distributions for the main parameters of the analysis with freely floating IGM temperature. The overall agreement between the previous results (green contours) and ours (blue and orange contours) is evident. Both have been obtained by using the same data set,~i.e.~MIKE/HIRES data, as described in Section~\ref{sec:data}. Firstly, we can notice the expected degeneracy between $T_0(z)$ and $\alpha$. Due to this degeneracy, different prior choices on the IGM thermal history may sensibly affect the limits on the WDM mass. That is why, by looking at the 1D distribution for $m_{\rm WDM}$ and $T_0(z=5.4)$, we can notice a discrepancy between our prediction (blue) and the one of Ref.~\cite{Irsic:2017ixq} (green). In the previous study, in fact, flat priors on $T_0(z)$ were assumed, only forbidding unphysical jumps $\Delta T_0 > 5000$~K between adjacent redshift bins, whereas the blue contours are obtained by imposing broad gaussian priors on $T_0(z)$ centred around their reference values, with standard deviation $\sigma = 3000$~K. The latter choice still constitutes a conservative assumption, given that it only prevents the temperatures to peak at $T_0 = 0$ at high redshifts, without precluding them from assuming reasonably cold values. 
Even though the blue 1D distribution is peaked at higher temperatures, only extremely cold values are excluded from the corresponding 2$\sigma$ region. Quantitatively, the previous analysis led indeed to a 2$\sigma$ lower limit on the thermal WDM mass $m_{\rm WDM} > 2.1~{\rm keV}$, while we have obtained $m_{\rm WDM} > 2.7$~keV.
For a more direct comparison with the previous results, we have also performed an analysis assuming flat priors on $T_0(z)$, shown by the orange contours in Fig.~\ref{fig:thermal_Tfree}. Under this assumption, which exactly corresponds to the prior choice adopted in Ref.~\cite{Irsic:2017ixq} for the freely floating temperature case, we have obtained $m_{\rm WDM} > 2.2$~keV (2$\sigma$), in excellent agreement with the previously published result.
Such agreement is also manifest when looking at the 1D posterior distribution for both $m_{\rm WDM}$ and $T_0(5.4)$. The orange curves are in very good agreement with the previous results, probing that the discrepancies between blue and green contours are driven by the different thermal history choice, rather than by some of the approximations characterising our interpolation scheme.

In Fig.~\ref{fig:thermal_Tpowlaw} we compare the 1D and 2D distributions for the main free parameters of the power law analysis, chosen to be the reference case in Ref.~\cite{Irsic:2017ixq} as well as in the present work. Analogously to the freely floating temperature plots, the green contours refer to the results of Ref.~\cite{Irsic:2017ixq}; the red contours represent our results. Most of the considerations that we have done above apply to this case too. The main difference between the results of the two analyses consists in a significant tightening of the upper bound on the thermal WDM mass, due to the less conservative prior choice, as it has been explained in Section~\ref{sec:method}. The previously published 2$\sigma$ limit is $m_{\rm WDM} > 4.1$~keV, whereas in the present work we have obtained a slightly weaker constraint, namely $m_{\rm WDM} > 3.6$~keV. In the aforementioned freely floating $T_0$ case, our study has yielded to a more aggressive limit with respect to Ref.~\cite{Irsic:2017ixq}. Conversely, in our reference analysis, the lack of cross simulations, which is the main difference characterising our work (see Section~\ref{sec:method}), has resulted in a weaker upper limit on the thermal WDM mass. From this point of view, the power law analysis that we have adopted as our reference MCMC analysis can be considered conservative.

\section{\label{app:bf} Best fit and confidence levels}

\begin{table}
  \setlength{\tabcolsep}{4pt}\renewcommand{\arraystretch}{1.2}
  \normalsize{
  \begin{tabular}{|c||c|c|c|} \hline
  \small{Parameter} & \small{(1$\sigma$)} & \small{(2$\sigma$)} & \small{Best Fit} \\ \hline\hline
$\bar{F}(z=4.2)$    & [0.35,~0.39]   & [0.33,~0.41]  &  0.35  \\  
$\bar{F}(z=4.6)$    & [0.27,~0.32]   & [0.25,~0.35]  &  0.26  \\  
$\bar{F}(z=5.0)$    & [0.16,~0.21]   & [0.15,~0.24]  &  0.18  \\  
$\bar{F}(z=5.4)$    & [0.05,~0.09]   & [0.02,~0.11]  &  0.07  \\  
$T_0^A~[10^{4}$~K]  & [0.55,~0.95]   & [0.41,~1.23]  &  0.74  \\  
$T_0^S$ 	        & [-5,~-2.72]    & [-5,~1.34]    &  -4.38  \\ 
$\widetilde{\gamma}^A$         & [1.35,~1.53]   & [1.21,~1.60]  &  1.45  \\  
$\widetilde{\gamma}^S$         & [-2.16,~-1.32] & [-2.41,~1.07] &  -1.93  \\ 
$\sigma_8$          & [0.67,~0.99]   & [0.53,~1.11]  &  0.84  \\  
$z_{\rm reio}$      & [7.73,~10.32]  & [7,~12.30]    &  9.16  \\  
$n_{\rm eff}$       & [-2.6,~-2.35]  & [-2.6,~-2.20] &  -2.46  \\ 
$f_{\rm UV}$        & [0,~1]	     &  [0,~1]       &  0.02  \\  
$\beta$             & [1.5,~10]      & [1.5,~10]     &  3.2  \\   
$\gamma$            & [-6.24,~-0.15] & [-10,~-0.15]  &  -4.8  \\  
$\alpha$~[Mpc$/h$]  & [0,~0.01]     & [0,~0.03]     &   0.005 \\ \hline
  \end{tabular}}
 \caption{\looseness=-1 Here we show the marginalised constraints at 1$\sigma$ and 2$\sigma$ C.L., as well as the best fit values for all the free parameters of our reference MCMC analysis (see the text for further details). Our best fit model has a $\chi^2/d.o.f.= 29/38$. \label{tab:bestfit}}
\end{table}

In this Appendix we report the full table of the best fit parameters and their 1$\sigma$ and 2$\sigma$ confidence intervals, corresponding to our reference MCMC analysis, namely to the assumption of a power law evolution for both the IGM temperature amplitude and slope (Table~\ref{tab:bestfit}).

By looking at Table~\ref{tab:bestfit}, one might notice that the lower limits on $\sigma_8$ and $n_{\rm eff}$ are sensibly underestimated with respect to the previous results.
This effect cannot be fully addressed to the overall weakening of the constraints due to the presence of two additional free parameters in the MCMC analysis.
As it was already mentioned in Section~\ref{sec:method}, this issue is partly due to intrinsic difficulties of the new interpolation scheme when sampling regions that are very far from the range covered by our simulations, a situation which can only occur for $\sigma_8$ and $n_{\rm eff}$. These two parameters are indeed the only ones for which we are scanning an interval of values spread significantly beyond the range covered by our simulations. Nevertheless, such parameters were nearly unconstrained even in the previous analyses, and our results are not biased by this problem.
We made sure of that by performing several times each of our analyses, for both the thermal case and the general one, imposing gaussian priors centred around Planck values for $\sigma_8$ and $n_{\rm eff}$~\cite{Ade:2015xua}, both individually and in combination, with various values for the standard deviations.
None of the runs provided sensibly different bounds on the other parameters. Thus, we can conclude that not being able to perfectly constrain extremely low values for $\sigma_8$ and $n_{\rm eff}$ does not affect any of our predictions.

\section{\label{app:validation} Robustness of the method}
\begin{figure}
\begin{minipage}{0.49\textwidth}
\vspace{0.2cm}
\subfloat[\label{fig:val1}]
{\includegraphics[width=\textwidth]{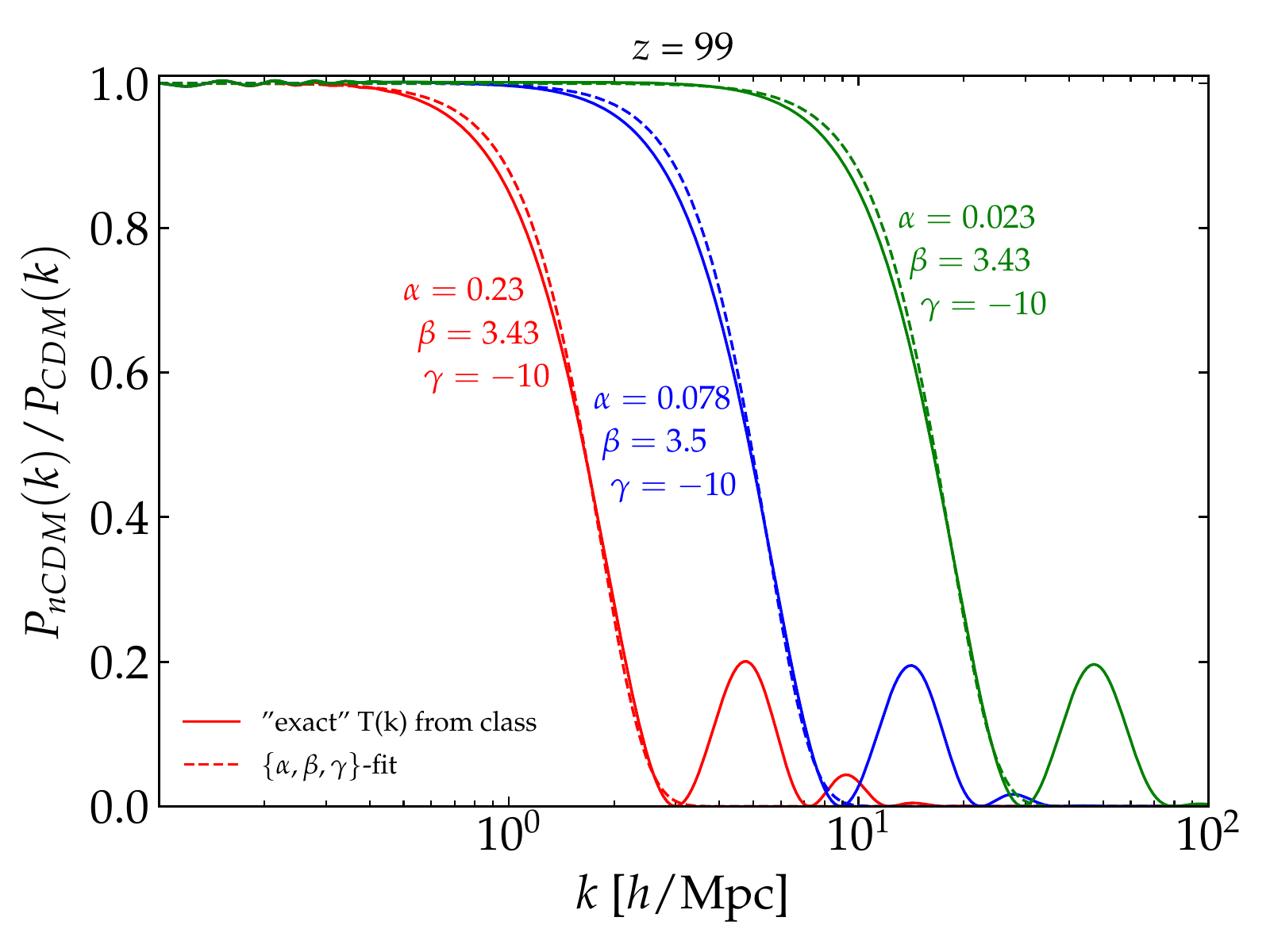}}
\vspace{0.5cm}
\end{minipage}
\begin{minipage}{0.48\textwidth}
\subfloat[\label{fig:val2}]
{\includegraphics[width=\textwidth]{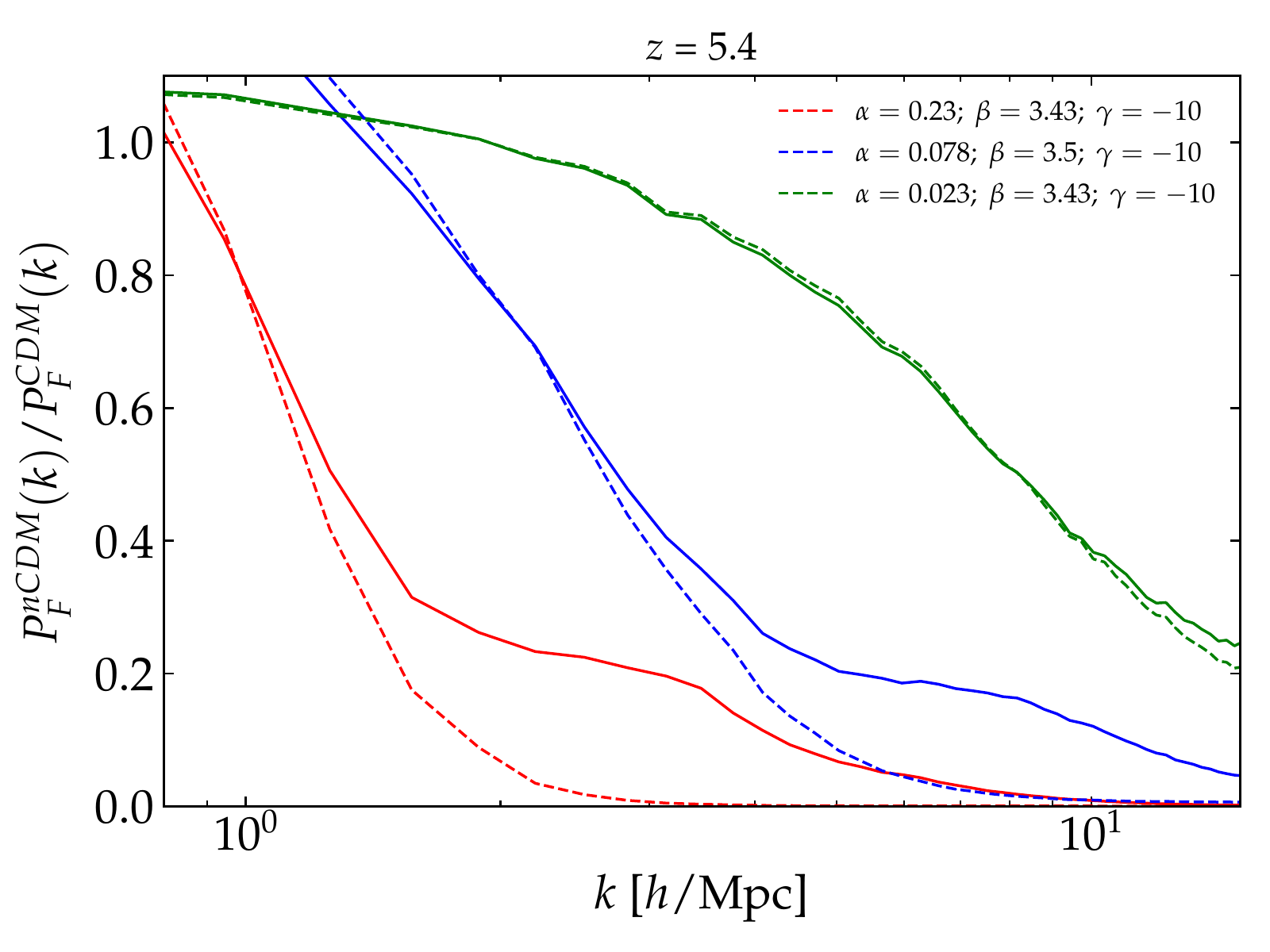}}
\vspace{0.1cm}
\end{minipage}
\caption{(a) In the upper panel we plot the "exact" squared transfer functions associated to three different ETHOS models (solid lines), all of them  featuring small-scale oscillations. We also plot the corresponding fitted transfer functions (dashed lines), obtained by neglecting such oscillations when fitting the "exact" ones.\\
(b) In the lower panel we plot the ratios with respect to a pure CDM model (at redshift $z=5.4$) for the flux power spectra extracted from two sets of simulations: the solid lines by using as initial conditions the "exact" transfer functions, the dashed lines by using as initial conditions the fitted $T(k)$.\\
The colour code of the two panels is the same.}\label{fig:val}
\end{figure}

\begin{figure*}
\includegraphics[width=0.9\textwidth]{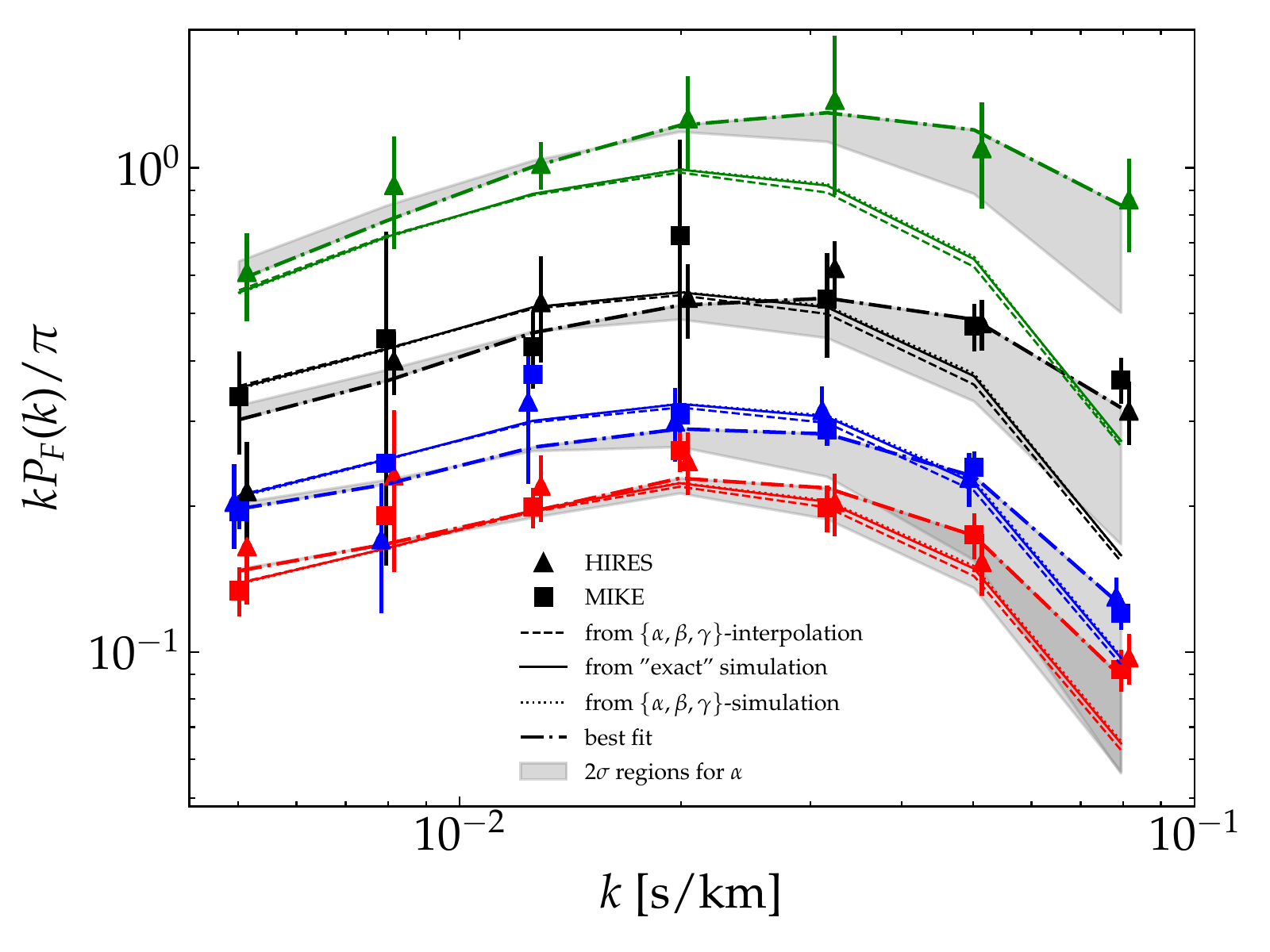}
\caption{Here we plot the flux power spectra for the most viable model shown in Fig.~\ref{fig:val}, together with MIKE/HIRES data points and error bars. Different colours stand for different redshifts,~i.e.,~from the bottom to the top,~$z=4.2,4.6,5.0,5.4$. See the text for the detailed discussion about the different power spectra listed in the legend.}
\label{fig:final}
\end{figure*}

This Appendix is devoted to test both the $\{\alpha,\beta,\gamma\}$-fitting procedure and the interpolation scheme illustrated in the previous Sections, in order to demonstrate the robustness and accuracy of our novel approach. To this end, we focus on three specific nCDM models belonging to the ETHOS class of models,~i.e.~associated to DM-Dark Radiation interaction scenarios with three different strengths. For our purposes, it is not necessary to go into the particle physics details of the models, we just need to notice that the corresponding transfer functions are characterised by a cut-off at the scales of interest for Lyman-$\alpha$ forest observations. Such squared transfer functions are plotted as solid lines in Fig.~\ref{fig:val1}, together with the corresponding $\{\alpha,\beta,\gamma\}$-fits (dashed lines).
The solid lines are dubbed as "exact" $T(k)$, since they have been produced by the numerical Boltzmann solver \texttt{class}~\cite{Lesg:2011}, where the non-standard interactions characterising the considered models are fully implemented. That is why they feature small-scale oscillations which our $\{\alpha,\beta,\gamma\}$-parametrisation is not able to capture.
For each squared transfer function plotted in Fig.~\ref{fig:val1}, we have also reported the corresponding $\{\alpha,\beta,\gamma\}$-values, obtained by fitting the solid curves down to their first minima, completely neglecting the oscillations.
In Fig.~\ref{fig:val2} we have plotted the ratios with respect to a pure CDM model (at redshift $z=5.4$) for the flux power spectra extracted from two sets of simulations: the solid lines by using as initial conditions the aforementioned "exact" transfer functions, the dashed lines by using as initial conditions the fitted ones. 
By analysing the two panels, for which we have adopted the same colour code, it is clear that differences between the "exact" flux power spectra and the $\{\alpha,\beta,\gamma\}$-predictions appear only when the power suppression with respect to the standard CDM case is more than 50$\%$. Furthermore, this is true only for nCDM models characterised by suppression at significantly large scales, which are indeed associated with $\alpha$-values well above the marginalised 2$\sigma$ upper limit obtained in the present work,~i.e.~$\alpha < 0.03$ Mpc/$h$. 
Let us highlight, in fact, that the fitted transfer function of the most viable model that we have considered (green dashed line), is practically superposed with the corresponding "exact" $T(k)$ (green solid line).
For both these reasons, we can conclude that the differences due to our inability to capture small-scale oscillations appear only for flux power spectra which lie anyway very far from the Lyman-$\alpha$ forest data points. This completely justifies ignoring such oscillations when applying our fitting procedure.

Let us now look at Fig.~\ref{fig:final}, where we have focused on the flux power spectra of the model described by the green curves in Fig.~\ref{fig:val}, given that its $\alpha$-value is the only one accepted at 2$\sigma$~C.L. by our analysis. Different colours stand for different redshifts,~i.e.,~from the bottom to the top, $z=4.2,4.6,5.0,5.4$. For each redshift bin we have plotted both MIKE and HIRES data points and error bars, as triangles and squares, respectively. Firstly we can note again the excellent agreement between the flux power obtained by using the "exact" initial conditions (solid lines) with respect to the fitted ones (dashed lines). Most importantly, the dotted lines correspond to flux power spectra computed by simply interpolating in our coarse grid, without running any dedicated simulation, and they nicely coincide with the solid ones, at each redshift.
It is thus evident that models associated with $\{\alpha,\beta,\gamma\}$-combinations sampling our grid of simulations are perfectly reproduced by our interpolation scheme.

As a reference, we have also plotted our best fit flux power spectra (dot-dashed lines), associated to the parameter values reported in Table~\ref{tab:bestfit}. The grey dashed areas represent instead the region spanned by flux power spectra with values of $\alpha$ varying up to its 2$\sigma$ marginalised upper bound.

\section{\label{app:area} Comparison with the Area Criterion}

\begin{figure}
\includegraphics[width=0.51\textwidth]{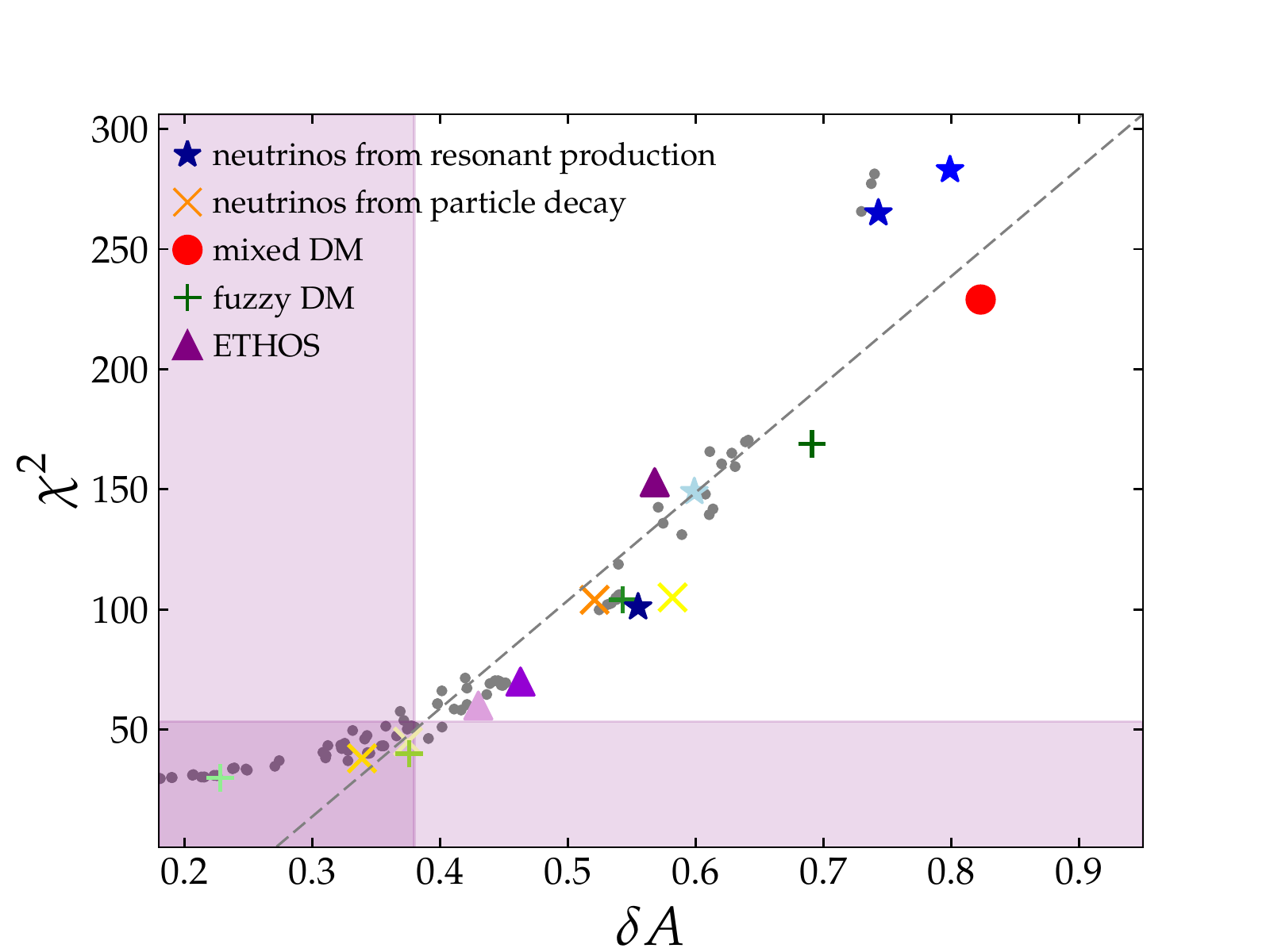}
\caption{Here we show the correlation between the area estimators $\delta A$ (from~Ref.~\cite{Murgia:2017lwo}) relative to each of the examined nCDM models, and the corresponding $\chi^2$ values, obtained with our reference MCMC analysis and reported in Table~\ref{tab:models}.
Different symbols refer to points belonging to different groups of nCDM models, consistently recalling the convention adopted in Fig.~\ref{fig:ABG}. The dashed line represents the linear regression fit to the results. The grey dots correspond to the $\{\alpha,\beta,\gamma\}$-combinations listed in Table~\ref{tab:params}, each of them associated to a different nCDM simulation. The vertical and horizontal shaded bands represent the regions corresponding to the 2$\sigma$ C.L. for the Area Criterion and the MCMC analysis, respectively. All those models sampling the lower left intersection between the two bands are thereby accepted (at 2$\sigma$~C.L.) by both the analyses, whereas models which sample the white region are rejected.}
\label{fig:area}
\end{figure}

In Ref.~\cite{Murgia:2017lwo} we introduced a simple method, based on linear perturbation theory, for testing different nCDM scenarios with Lyman-$\alpha$ forest data by using an approximate yet intuitive estimator, and we named it as \emph{Area Criterion}. With such method it is not possible to extract absolute limits on the nCDM parameters, but it allows to look into deviations with respect to a given reference case, which is typically chosen to be the most updated bound from a full statistical analysis~\cite{Murgia:2017lwo,Murgia:2017cvj,takeshi}. We address the reader to any of the quoted references for further details. However, let us briefly recall how the deviation of a given nCDM model with respect to the standard CDM case is parametrised,~i.e.~by the ratio
\begin{equation}\label{eq:rk}
 \xi(k) = \frac{P_{1\rm{D}}(k)}{P^{\rm{CDM}}_{1\rm{D}}(k)},
\end{equation}
where $P_{1\rm{D}}(k)$ is the 1D power spectrum of the model that we are considering, computed through the following integral on the 3D linear matter power spectrum, $P(k')$:
\begin{equation}\label{eq:pk1d}
 P_{1\rm{D}}(k)=\frac{1}{2\pi} \int\limits_k^\infty {\rm d}k'k'P(k').
\end{equation}
The suppression in the power spectra is then computed via the following estimator:
\begin{equation}\label{eq:deltaA}
 \delta A \equiv \frac{A_{\rm CDM} - A}{A_{\rm CDM}},
\end{equation}
\looseness=-1 where $A$ is the integral of $\xi(k)$ over the range of scales probed by Lyman-$\alpha$ observations,~i.e.
\begin{equation}\label{eq:A}
A = \int\limits_{k_{\rm min}}^{k_{\rm max}} {\rm d}k\ \xi(k),
\end{equation}
so that $A_{\rm CDM} \equiv k_{\rm max} - k_{\rm min}$, by construction.

Notice that the choice of the reference thermal WDM model for calibrating the Area Criterion is crucial for establishing the threshold which defines which models to accept/reject. In our previous work, we performed an analysis in the $\{\alpha,\beta,\gamma\}$-space by calibrating the method with $m_{\rm WDM} = 3.5$~keV (i.e.,~the 2$\sigma$ limit from the conservative MIKE/HIRES+XQ-100 analysis of Ref.~\cite{Irsic:2017ixq}). We quantified the reference power suppression through the area estimator $\delta A$, and we rejected (at 2$\sigma$ C.L.) all those nCDM models that feature a larger power suppression with respect to the reference one. Since the aforementioned 2$\sigma$ reference limit roughly coincides with the constraint obtained in the present work for thermal WDM masses (power law analysis), we can consistently compare the results presented here with the approximate conclusions reported in Ref.~\cite{Murgia:2017lwo}. It is interesting, indeed, to quantify the precision of the simple Area Criterion with respect to the full statistical data analysis that we have illustrated in this work.

Firstly, let us note that by marginalising the Area Criterion results over $\beta$ and $\gamma$, we obtained the following upper limit on $\alpha < 0.058~{\rm{Mpc}}/h$~(2$\sigma$)~\cite{Murgia:2017lwo,Murgia:2017cvj}, which is weaker with respect to the bounds quoted in Section~\ref{sec:results} of this work. This could mostly be due to the unavoidably prominent tail at large values in the 1D $\alpha$-distribution, when the analysis is done with the approximate area method. As we have pointed out in this work, such tail corresponds to extreme values of both $\beta$ and $\gamma$, as well as very cold IGM temperatures, unlikely to be physically motivated. By simply applying the Area Criterion it is intrinsically impossible to account for this aspect.

Let us now compare the results listed in Table~\ref{tab:models} with the conclusions reported in Table 4 of Ref.~\cite{Murgia:2017lwo}, which have been obtained by applying the Area Criterion to the same nCDM particle model examples analysed in this work.
In Fig.~\ref{fig:area} we show the correlation between the area estimators $\delta A$ relative to each of the examined nCDM models and the corresponding $\chi^2$ values, obtained with our reference MCMC analysis and reported in Table~\ref{tab:models}.
The two different sets of predictions are visibly correlated, with correlation coefficient $r = 0.94$. The dashed line represents the linear regression fit to the results. Different symbols are used for identifying points belonging to different groups of nCDM models, consistently recalling the convention adopted in Fig.~\ref{fig:ABG}. The vertical and horizontal shaded bands represent the regions which are included at 2$\sigma$ C.L. by the Area Criterion and the MCMC analysis, respectively. All those models sampling the lower left intersection between the two bands are thus accepted (at 2$\sigma$ C.L.) by both the analyses, whereas models which sample the white region are excluded by both of them.
Interestingly, none of the particle model examples examined is rejected by one analysis and accepted by the other. 

When considering also the grey dots, which correspond to the $\{\alpha,\beta,\gamma\}$-triplets listed in Table~\ref{tab:params}, the correlation between the conclusions drawn by the two methods is even more evident. Since the grey dots which sample the vertical shaded band refer to models accepted at 2$\sigma$ C.L., they are also associated to the red transfer functions shown in Fig.~\ref{fig:Tk2s}. Note that the departure from the linear correlation occurring for small values of $\delta A$ is an intrinsic feature of the method. Whereas the $\chi^2$ distribution saturates when approaching the best fit $\chi^2$ value, the area estimator can assume arbitrarily small (positive) values. It is remarkable that, among 109 models thoroughly sampling the $\{\alpha,\beta,\gamma\}$-space, only two of them are accepted by the Area Criterion while rejected by the MCMC analysis. Conversely, it is worthwhile to notice that only one borderline $\{\alpha,\beta,\gamma\}$-triplet is rejected by the Area Criterion while accepted by the MCMC analysis, that is a confirmation of the suitability of the former as an approximate yet effective and conservative method.
Therefore, the intuitive Area Criterion seems to be a very good approximation for performing preliminary tests on non-standard DM scenarios in an immediate and simplified way.

\acknowledgments{MV and RM are supported by the INFN INDARK PD51 grant. VI is supported by the US NSF grant AST-1514734. RM acknowledges the hospitality of RWTH Aachen University where the project was finalised. The simulations were performed on the Ulysses SISSA/ICTP supercomputer. RM is thankful to Elias S. Kammoun and Andrej Obuljen for useful discussions.}

\bibliography{wdm}
\bibliographystyle{unsrt}

\end{document}